\documentclass[sigplan, screen, authorversion]{acmart}

\copyrightyear{2022} 
\acmYear{2022} 
\setcopyright{acmlicensed}\acmConference[PLDI '22]{Proceedings of the 43rd ACM SIGPLAN International Conference on Programming Language Design and Implementation}{June 13--17, 2022}{San Diego, CA, USA}
\acmBooktitle{Proceedings of the 43rd ACM SIGPLAN International Conference on Programming Language Design and Implementation (PLDI '22), June 13--17, 2022, San Diego, CA, USA}
\acmPrice{15.00}
\acmDOI{10.1145/3519939.3523731}
\acmISBN{978-1-4503-9265-5/22/06}

\usepackage{siunitx}

\usepackage{microtype} 

\usepackage{booktabs}   
\usepackage{subcaption} 

\usepackage{longtable}

\usepackage{enumitem}

\usepackage{listings}
\newcommand{\norm}[2][R]{
  \ensuremath{
    {{#2}\!\downarrow_{#1}}
  }
}

\newcommand{\rlnm}[1]{
  \ensuremath{
    (\textsf{#1})
  }
}

\usepackage{tikz}
\usetikzlibrary{tikzmark}
\usetikzlibrary{arrows.meta}
\tikzset{
  trace/.style={line width=1ex, line cap=round, opacity=0.3, rounded corners, transform canvas={xshift = -0.5ex, yshift = 0.5ex}},
  progress/.style={fill, circle, radius=1.5ex}
}

\usepackage{prftree}

\usepackage{savesym}
\savesymbol{comment}
\usepackage[nompar, nosign]{commenting}
\declareauthor{sr}{Steven}{blue}
\authorcommand{sr}{comment}
\declareauthor{ej}{Eddie}{red}
\authorcommand{ej}{comment}
\declareauthor{lo}{Luke}{cyan}
\authorcommand{lo}{comment}

\usepackage{apxproof}

\usepackage{mathtools} 
\newcommand\defeq{\coloneqq}

\usepackage[capitalise]{cleveref} 

\newcommand\defn[1]{{\bf \emph{#1}}}

\usepackage{amsthm}
\theoremstyle{definition} 
\newtheorem{definition}{Definition}[section]
\newtheorem{example}[definition]{Example}

\theoremstyle{remark} 
\newtheorem{remark}{Remark}[section]

\theoremstyle{plain} 
\newtheorem{theorem}{Theorem}[section]
\newtheorem{lemma}[theorem]{Lemma}
\newtheorem{corollary}[theorem]{Corollary}

\begin{abstract}
We propose a new cyclic proof system for automated, equational reasoning about the behaviour of pure functional programs. 
The key to the system is the way in which cyclic proofs and equational reasoning are mediated by the use of contextual substitution as a cut rule.
We show that our system, although simple, already subsumes several of the approaches to implicit induction variously known as ``inductionless induction'', ``rewriting induction'', and ``proof by consistency''.
By restricting the form of the traces, we show that global correctness in our system can be verified incrementally, taking advantage of the well-known size-change principle, which leads to an efficient implementation of proof search.
Our CycleQ tool, implemented as a GHC plugin, shows promising results on a number of standard benchmarks.
\end{abstract}

\begin{document}

\ccsdesc[500]{Theory of computation~Equational logic and rewriting}
\ccsdesc[500]{Theory of computation~Logic and verification}

\keywords{Cyclic proofs, equational reasoning, rewriting induction, inductionless induction.}  

\title[{CycleQ:\ An Efficient Basis for Cyclic Equational Reasoning}]{CycleQ:\ An Efficient Basis\\ for Cyclic Equational Reasoning}

\author{Eddie Jones}
\affiliation{
  \department{Department of Computer Science}
  \institution{University of Bristol}
  \city{Bristol}
  \country{United Kingdom}
}

\author{C.-H. Luke Ong}
\affiliation{
  \department{Department of Computer Science}
  \institution{University of Oxford}
  \city{Oxford}
  \country{United Kingdom}
}

\author{Steven Ramsay}
\affiliation{
  \department{Department of Computer Science}
  \institution{University of Bristol}
  \city{Bristol}
  \country{United Kingdom}
}

\maketitle

\newcommand{\mapE}{\mathtt{mapE}}
\newcommand{\mapT}{\mathtt{mapT}}
\newcommand{\cstK}{\mathtt{Cst}}
\newcommand{\varK}{\mathtt{Var}}
\newcommand{\appK}{\mathtt{App}}
\newcommand{\mkE}{\mathtt{MkE}}

\begin{figure*}[ht]
  \begin{minipage}{\textwidth}
    \[
      \begin{array}{c}
        \prftree{
          \prftree{
            \prftree{
              \prftree{n \doteq n}
            }
            {
              \prftree{
                \prftree{
                  \prftree{\varK\ v \doteq \varK\ v}
                }
                {
                  \mapT\ \mathtt{id}\ (\varK\ v) \doteq (\varK\ v)
                }
              }
              {
                \prftree{
                  \prftree{\cstK\ c \doteq \cstK\ c}
                }
                {
                  \mapT\ \mathtt{id}\ (\cstK\ c) \doteq (\cstK\ c)
                }
              }
              {
                \prftree{
                  \prftree{
                    \prftree{
                      \textrm{(0)}
                    }
                    {
                      \prftree{e_1 \doteq e_1}
                    }
                    {
                      \mapE\ \mathtt{id}\ e_1 \doteq e_1
                    }
                  }
                  {
                    \prftree{
                      \textrm{(0)}
                    }
                    {
                      \prftree{e_2 \doteq e_2}
                    }
                    {
                      \mapE\ \mathtt{id}\ e_2 \doteq e_2
                    }
                  }
                  {
                    \appK\ (\mapE\ \mathtt{id}\ e_1)\ (\mapE\ \mathtt{id}\ e_2) \doteq \appK\ e_1\ e_2
                  }
                }
                {
                  \mapT\ \mathtt{id}\ (\appK\ e_1\ e_2) \doteq \appK\ e_1\ e_2
                }
              }
              {
                \mapT\ \mathtt{id}\ t \doteq t
              }
            }
            {
              \mkE\ (\mapT\ \mathtt{id}\ t)\ n \doteq \mkE\ t\ n
            }
          }
          {
            \mapE\ \mathtt{id}\ (\mkE\ t\ n) \doteq \mkE\ t\ n
          }
        }
        {
          \textrm{0: } \mapE\ \mathtt{id}\ e \doteq e
        }
      \end{array}
    \]
  \end{minipage}
  \caption{A cyclic proof of $\mathtt{mapE}\ \mathtt{id}\ e \doteq e$.}\label{fig:map-id-e-prf}
\end{figure*}

\section{Introduction}

An advantage of pure functional programming is the ease with which one can reason about the behaviour of programs.
Interesting properties can often be proven using only a combination of induction and equational reasoning.

However, as is well known, inductive theorem proving is challenging.
The incompleteness of typical inductive theories and the non-analyticity of their induction rules excludes a general algorithmic solution~\cite{bundy1999automation}.
Moreover, even if we restrict our attention to what we might loosely imagine are the cases we care about, namely those functional programs that occur in practice, the situation is still extremely complex.

Functional programmers employ a variety of inductive and mutually inductive datatypes and rarely restrict themselves to functions defined by structured recursion schemes.
Hence, not only do we need induction principles for each datatype, but we should expect that these schemes can be nested, combined for mutual induction, or generalised to account for non-structural recursion.

Despite this, there are already several tools that have shown success at automatically proving equational properties of functional programs.
However, to the best of our knowledge, none have a smooth treatment of the more complicated induction schemes that are frequently required in practice.
For example, proofs that require mutual induction are not supported by default in either HipSpec~\cite{claessen2013automating}, IsaPlanner~\cite{dixon2003isaplanner} or Zeno~\cite{sonnex2012zeno}, and reasoning about mutually recursive functions is described in the ACL2 manual as being ``a bit awkward''~\cite{acl2-manual}.
Moreover, in several of these tools, mutually inductive datatypes are simply not supported.

Using induction schemes that are tailored to specific conjectures is important; although automatic lemma discovery techniques can sometimes compensate,
they have a number of weaknesses such as limited applicability, over generalisation, and scalability for complex formulas~\cite{johansson2019lemma}.
Hence, although lemma discovery is crucial in all but the simplest inductive proofs, any improvement to the underlying inductive proof system, reducing the burden on lemma generation heuristics, is worthwhile.

In this paper, we propose a novel \emph{cyclic proof system for equational reasoning} and an accompanying algorithm for efficient proof search, which we have implemented as a plugin for GHC --- CycleQ.
The system seamlessly supports complex forms of inductive argument, such as nested or mutual induction, and is agnostic about lemma discovery techniques (which we leave aside as an orthogonal concern).

\subsection{Cyclic Proofs and Equational Reasoning}

Cyclic proofs occupy a part of non-wellfounded proof theory, in which infinite proof trees are required to be regular, e.g. representable as a finite graph~\cite{brotherston2005cyclic,sprenger2003}.
Unsound arguments are excluded by requiring a \emph{global correctness condition} on the infinite paths, such as inclusion in a particular \( \omega \)-regular language.
The regularity restriction makes cyclic proofs quite well behaved, and there has recently been a number of works exploring the theoretical and practical advantages of this form of circular reasoning~\cite{tsukada2021software,brotherston2005cyclic,brotherston2008cyclic,brotherston2012generic,stratulat2019efficient,kuperberg2021cyclic,stratulat2021cyclist,itzhaky2021cyclic,das2018,das2019,das2021}.

On the practical side, one of the major driving forces has been the potential to improve state-of-the-art automated reasoning.
Cyclic proof systems appear to better capture the exploratory nature of goal-directed proof search, especially with respect to ``inductive'' reasoning.
A particular advantage is the ability to avoid committing to either a fixed menu of  induction schemes or a fixed choice of induction variables in advance.
Rather, systems can justify a circular argument post hoc through an appeal to infinite descent bespoke to the proof structure discovered by the search.

For example, consider a mutually inductive definition of two types comprising annotated syntax trees\footnote{A typical annotation is the line and column numbers marking their provenance in some source code, but here we use a single natural number for simplicity.} in Haskell:
\begin{lstlisting}
  data Term a         data Expr a
    = Var a             = MkE (Term a) Nat
    | Cst Nat
    | App (Expr a) (Expr a)
\end{lstlisting}

We can define two functions: \( \mathtt{mapT} \) and \( \mathtt{mapE} \), that express the functoriality of these type constructors and conjecture that the relevant laws hold.
\cref{fig:map-id-e-prf} shows the cyclic proof obtained by our system for the identity law for \( \mapE \).
Here, and elsewhere, the cycle is presented by labelling a node in the proof tree with a number, e.g. labelling the root \( 0 \), and using this label elsewhere as a premise without further justification.
Equations are given using the symbol \( \doteq \) to emphasize that they are regarded as unordered (i.e.\ the left- and right-hand sides are interchangeable).
See \cref{ex:Unsound preproof} and the following remark for a fuller description of this notation.

\begin{figure*}[ht]
  \begin{minipage}{1\textwidth}
    \[
      \begin{array}{c}
        \prftree
        {
          \prftree
          {
            \mathtt{Nil} \doteq \mathtt{Nil}
          }
        }
        {
          \prftree
          {\prftree{y \doteq y}}
          {
            \prftree
            {(\textrm{0})}
            {
              \prftree
              {\mathtt{take}\ (\mathtt{len}\ zs)\ (\mathtt{Cons}\ z\ zs) \doteq \mathtt{take}\ (\mathtt{len}\ zs)\ (\mathtt{Cons}\ z\ zs)}
            }
            {
              \mathtt{butLast}\ (\mathtt{Cons}\ z\ zs) \doteq \mathtt{take}\ (\mathtt{len}\ zs)\ (\mathtt{Cons}\ z\ zs)
            }
          }
          {
            \mathtt{Cons}\ y\ (\mathtt{butLast}\ (\mathtt{Cons}\ z\ zs)) \doteq \mathtt{Cons}\ y\ (\mathtt{take}\ (\mathtt{len}\ zs)\ (\mathtt{Cons}\ z\ zs))
          }
        }
        {
          \textrm{0: } \mathtt{butLast}\ (\mathtt{Cons}\ y\ ys) \doteq \mathtt{take}\ (\mathtt{len}\ ys)\ (\mathtt{Cons}\ y\ ys)
        }
        \\[4mm]

        \prftree{
          \prftree
          {
            \mathtt{Nil} \doteq \mathtt{Nil}
          }
        }
        {
          \prftree
          {
            \prftree
            {
              \mathtt{Nil} \doteq \mathtt{Nil}
            }
          }
          {
            \prftree
            {\prftree{x \doteq x}}
            {(\textrm{0})}
            {
              \mathtt{Cons}\ x\ (\mathtt{butLast}\ (\mathtt{Cons}\ y\ ys)) \doteq \mathtt{Cons}\ x\ (\mathtt{take}\ (\mathtt{len}\ ys)\ (\mathtt{Cons}\ y\ ys))
            }
          }
          {
            \mathtt{butLast}\ (\mathtt{Cons}\ x\ xs) \doteq \mathtt{take}\ (\mathtt{len}\ xs)\ (\mathtt{Cons}\ x\ xs)
          }
        }
        {
          \mathtt{butLast}\ xs \doteq \mathtt{take}\ (\mathtt{len}\ xs - \mathtt{S}\ \mathtt{Z})\ xs
        }
      \end{array}
    \]
  \end{minipage}
  \Description{}
  \caption{A cyclic proof of \( \mathtt{butLast}\ xs \doteq \mathtt{take}\ (\mathtt{len}\ xs - \mathtt{S}\ \mathtt{Z}) \doteq xs \).}\label{fig:butLast-proof}
\end{figure*}

Without a proper treatment of mutual induction, an inductive theorem prover would have to guess, heuristically, a strengthening of the inductive property, e.g.\ adding the conjunct \( \mapT\ \mathtt{id}\ t \doteq t \) to the original goal.
In our cyclic system, however, the two cycles depicted using label 0 fall out naturally from equational reasoning, and the fact that each involves a decrease, i.e.\ that the proof is globally correct, is easily verified.

A key consideration in the design of an automated cyclic proof system is how to control the formation of cycles in the proof.
Technically, it is sound to form cycles whenever proof search discovers a node of the proof tree that is logically stronger than an ancestor and for which the newly formed cycle would satisfy the global correctness condition.
Since, in general, we cannot expect there to be any syntactical relationship between the node and its ancestor, the formation of cycles is closely related to the use of cuts in the proof. Indeed \citet{tsukada2021software} have demonstrated that many techniques developed for efficient software model checking can be viewed as the introduction of cuts into cyclic proofs to discharge proof obligations earlier.


In Brotherston, Gorogiannis and Petersen's state-of-the-art \textsc{Cyclist}\(_\textsc{FO}\) prover, cycles are formed by a restricted kind of cut in which the node follows from its ancestor by a combination of weakening and instantiation~\cite{brotherston2012generic}.
However, the authors note that the lack of a more general cut rule and the lack of native support for equational reasoning causes their system to have difficulty with heavily-equational goals, such as  the commutativity of addition: \( x + y = y + x \).  They conjecture that a cyclic proof could be obtained if the lemma \(x + \mathtt{S}\ y = \mathtt{S}\ (x + y) \) were supplied as a hint.


In fact, the cyclic proof system that we develop in this paper can prove the commutativity of addition automatically, without any externally supplied lemma such as the one above.
The synthesized proof is given in \cref{fig:Commutative addition}, but we defer its discussion until later.

Our system consists of four rules: the reflexivity of equality, evaluation of program expressions, reasoning by cases, and substitution of equals for equals.
The key is the way that cyclic proof and equational reasoning are mediated through the use of (contextual) substitution as a cut rule.
\[
  \prftree[l]
  {\rlnm{Subst}}
  {M \doteq N}
  {C[N\theta] \doteq P}
  {C[M\theta] \doteq P}
\]
We refer to the left-hand premise of this rule as \emph{the lemma} and the right-hand premise as \emph{the continuation}.
This rule says that given a lemma \( M \doteq N \) and a goal \( C[M\theta] \doteq P \) containing an instance of \( M \), the proof can be continued by solving \( C[N\theta] \doteq P \) in which the instance of \( M \) has been replaced by a matching instance of \( N \).

Although, in principle, the choice of equation comprising the lemma may be completely unrelated to the rest of the proof tree (e.g.\ it may be supplied by a human or conjectured by a theory exploration tool), our proof search algorithm is able to synthesize proofs for 61\% of the relevant problems from the IsaPlanner benchmark suite whilst only choosing lemmas \( M \doteq N \) that already occur as nodes within the same proof tree, i.e.\ without the need to invoke any potentially costly lemma discovery technology.
For example, our system can prove \( \mathtt{butLast}\ xs \doteq \mathtt{take}\ (\mathtt{len}\ xs - \mathtt{S}\ \mathtt{Z}) \doteq xs \) in \textasciitilde\SI{40}{ms}.
A proof can be seen in \cref{fig:butLast-proof}.
By comparison, HipSpec fails to prove the same result after \textasciitilde\SI{40}{s}, an attempt that involved 22 synthesised lemmas, 12 of which failed~\cite{rosenHipspecEval}.

The substitution rule can be seen in the two rule applications of \cref{fig:map-id-e-prf} that have (0) as a premise.
Here, the lemma is chosen to be the node labelled 0 at the root of the tree and the continuation is simply discharged by reflexivity.
The usage in \cref{fig:butLast-proof} is similar.
In the proof of the commutativity of addition \cref{fig:Commutative addition}, the continuation labelled (2) is much more complex and contains a nested inductive argument.

\subsection{Simulation of Inductionless Induction}

It is well known that cyclic proof systems can already simulate explicit structural induction schemes, and we additionally show that our system subsumes various kinds of implicit induction based on Knuth-Bendix completion, such as ``inductionless induction'' and ``proof by consistency'', that were intensively studied in the 1980s and 1990s, e.g.~\cite{musser1980proving,huet1982proofs,kapur1987proof,bundgen1989,fribourg1986,dershowitz1982,Jouannaud1989,kapur1986}.

On the surface, these approaches seem quite distinct from cyclic proofs; rather than proving a conjecture by induction, they posit it as an axiom and attempt to show that the resulting theory is consistent.
In order to connect the approaches, we use \emph{term rewriting induction}, in the sense of \citet{reddy1990term}, as a stepping stone, which is already known to subsume proof by consistency.
The key is to observe that the unconstrained use of hypotheses in Reddy's system gives rise to the structure of cyclic preproofs and that global correctness is guaranteed by construction as progress proceeds by rewriting and equations are orientated by a fixed (well-founded) reduction order.

Term rewriting approaches to induction share some of the advantages of cyclic proofs.
They support mutual induction, for example, and do not require a fixed induction scheme in advance.
However, our analysis also highlights a disadvantage by comparison with our cyclic system: rewriting approaches require that any equations discovered by the proof search be orientable with respect to the fixed reduction order.
Systems are not only very sensitive to the choice of the order in practice, but this requirement also precludes theorems like the above commutativity of addition, the symmetry of which is inherently unorientable.
For a critique see the 1988 POPL paper of \citet{garland1988}.

\subsection{The CycleQ Theorem Prover}

Since our proof system is quite simple, it is straightforwardly amenable to a goal-directed proof-search algorithm.
However, a na\"ive implementation will quickly run into performance difficulties.

One source is in the number of nodes that are candidates for cycles.
As mentioned previously, the formation of cycles is enabled by the \rlnm{Subst} rule restricted to employ only existing nodes of the current proof tree as the lemma $M \doteq N$.
However, the number of eligible lemmas to consider will consequently grow with the size of the proof.

The number of eligible lemmas can be drastically reduced by using a number of further restrictions motivated by redundancies we identify in the structure of proofs.
For example, if a lemma is itself justified by the \rlnm{Subst} rule, we can use its premise directly as contexts and substitutions are composable.

Another bottleneck is in the verification of the global correctness condition.
This source of inefficiency was already identified for the \textsc{Cyclist} prover, where a large proportion of the overall proof time is spent checking the global correctness of proof trees that turn out to be unsound~\cite{stratulat2019efficient}.
In this work, we avoid a similar problem by restricting our attention to variable-based traces and exploiting the incremental nature of goal-directed proof search.
We annotate the proof graph with an abstract domain representing the \(\omega \)-regular language of paths --- Lee, Jones and Ben-Amram's size-change graphs~\cite{lee2001size}.
Encoding the information directly in the proof graph allows the global correctness argument to be updated as each node is uncovered, and thus, unsound cycles are represented explicitly so that proof search can be terminated as soon as one is detected.
Furthermore, there is no recomputation of soundness for shared proof fragments.

We implemented our approach as a plugin for GHC called CycleQ.
It currently supports a small subset of Haskell, including top-level recursive functions, algebraic datatypes, and polymorphism.
Our evaluation on a number of benchmarks shows that it performs well on standard and mutual induction problems and can sometimes prove more complex goals that would typically require lemmas in other systems.

\paragraph{Contributions}

Our main contributions are as follows:
\begin{itemize}[leftmargin=*]
  \item We identify contextual substitution as the appropriate means for introducing cycles into an equational proof, presenting a simple proof system based on this mechanism.
  \item We show that, when targeting proofs about functional programs, our system subsumes approaches to implicit induction, known variously as ``inductionless induction'', ``proof by consistency'' and ``rewriting induction''.
  \item We show that, by restricting to variable traces, the global correctness condition of cyclic proof reduces to Lee, Jones and Ben-Amram's size-change principle.
        This approach leads directly to an efficient and \emph{incremental} procedure for detecting and verifying cycles based on size-change graphs.
  \item We identify several natural restrictions on contextual substitution that allow it to play the role of an efficient matching function for detecting potential cycles.
        Our evaluation on a number of benchmarks shows that it performs well on standard and mutual induction problems and can sometimes prove more complex goals that would typically require lemmas in other systems.
\end{itemize}

\paragraph{Outline}

The remainder of the paper is structured as follows.
In Section~\ref{sec:prelim}, we introduce necessary preliminaries and, in Section~\ref{sec:cyclic-proofs}, we present our simple cyclic proof system for equational reasoning.
In Section~\ref{sec:Rewriting induction}, we show that the system already subsumes Reddy's system of rewriting induction.
In Section~\ref{sec:Algorithmic system}, we develop the heuristics necessary for making the formation of cycles efficient, and we show that size-change termination can be used to enable an incremental approach to checking the global correctness condition.
A description of our implementation and its evaluation comprise Section~\ref{sec:Implementation}, and we conclude in Section~\ref{sec:Related work} with a discussion of related work.

\section{Preliminaries}\label{sec:prelim}

For the purpose of this formalism, we will consider a higher-order rewriting system and its induced equational theory.
Although the intended application of our work is functional programs, this setting is more general and facilitates direct comparison with rewriting induction (\cref{sec:Rewriting induction}).

In this section, we will cover some definitions from term rewriting used throughout the paper.

We assume a fixed \defn{signature} consisting of a finite set of algebraic datatypes \( D \) and function symbols \( \Sigma \).

For the types of our formal system, we use simple types built over \( D \), i.e. \( \tau,\, \sigma \defeq d \in D \mid \tau \rightarrow \sigma \).
The \emph{order} of a type is defined as follows:
\begin{align*}
  \textsf{ord}(d)                       & \defeq 0                                                         \\
  \textsf{ord}(\tau \rightarrow \sigma) & \defeq \max \{ \textsf{ord}(\tau) + 1,\, \textsf{ord}(\sigma) \}
\end{align*}

Each function symbol is assigned a type, written \( f : \tau \in \Sigma \).
Furthermore, function symbols are partitioned into a set of \defn{constructors} \( \Sigma_\textrm{con} \) (e.g. $\mathtt{Cons}$, $\mathtt{Nil}$, $\mathtt{Zero}$, $\mathtt{Succ}$), which are required to be at most first-order, and \defn{defined functions} \( \Sigma_\textrm{def} \) (e.g. $\mathtt{map}$, $\mathtt{add}$).
We write \( \Sigma_\textrm{con}(d) \) for the set of constructors whose 
return type is \( d \).






\defn{Terms} are generated from application, function symbols \( \Sigma \), and variables drawn from some countable set.
\[
  M,\, N \Coloneqq x \mid f \in \Sigma \mid M\ N
\]
As usual, we associate applications to the left.

A \defn{type environment}, typically \( \Gamma \) or \( \Delta \), is a set of variable-type pairs, written \( x : \tau \).
We will write \( \Gamma,\, \Delta \) (or \( \Gamma,\, x_0 : \tau_0,\, \dots,\, x_n : \tau_n \)) for the disjoint union of two environments.
The judgement \( \Gamma \vdash M : \tau \), defined by usual typing rules, asserts that \( M \) is a well-typed term of type \( \tau \) for the environment \( \Gamma \).

In what follows, we will restrict our attention to well-typed terms, but we omit the rules for simple typing, which are standard.




Terms give rise to a natural set of \defn{(one-hole) contexts}, generically written \( C[\cdot] \), which we define as:
\[
  C[\cdot] \Coloneqq \cdot \mid C[\cdot]\ M \mid M \ C[\cdot]
\]
where \( M \) ranges over terms.
We write \( C \circ D \) for their composition, where \( (C \circ D)[X] \defeq C[D[X]] \) for all terms \( X \).

A term \( M \) is \defn{subterm} of \( N \), written \( M \unlhd N \), if there exists a context \( C \) such that \( C[M] = N \).
When the witness \( C \) is non-trivial, i.e.\ not \( \cdot \), we write \( M \lhd N \).

\begin{lemma}\label{lem:Subterm order}
  \( {\unlhd} \) is a well-founded, partial order.
  \begin{appendixproof}
    \phantom{-}
    \paragraph{Reflexivity}
    Clearly, \( M \unlhd M \) witnessed by the trivial context.

    \paragraph{Antisymmetry}
    Suppose \( M \unlhd N \) and \( N \unlhd M \).
    That is, \( C[M] = N \) and \( D[N] = M \).
    In particularly, we have that \( N = C[D[N]] \).
    It follows that both \( C \) and \( D \) are trivial.
    Thus, \( M = N \) as required.

    \paragraph{Transitivity}
    If \( M \unlhd N \) and \( N \unlhd L \), we have that \( C[M] = N \) and \( D[N] = L \) thus \( D[C[M]] = L \).
    Therefore, \( M \unlhd L \) as required.

    \paragraph{Well-foundedness}
    Let \( N \) be some term.
    We shall show by induction that there are only finitely many terms such that \( M \unlhd N \):
    \begin{itemize}
      \item If \( N \) is a variable \( x \), then clearly the only term \( M \) such that \( C[M] = x \) for some \( C[\cdot] \) is \( x \).
      \item If \( N \) is a function symbol, the argument is analogous.
      \item If \( N \) is an application \( R\ L \). Suppose \( C[M] = N \) for some term \( M \). We shall show that \( M \) is from a finite set by case analysis of \( C[\cdot] \):
            \begin{itemize}
              \item If is trivial, then \( M = N \).
              \item If \( C[\cdot] \) is of the form \( R\ C'[\cdot] \) and, by induction, \( M \) is from a finite set.
              \item Or it is of the form \( C'[\cdot]\ L \) and the same argument applies.
            \end{itemize}
            As the finite union of finite sets is also finite, there are only finitely many such \( M \).
    \end{itemize}
    It immedaitely follows that \( \unlhd \) is well-founded.
  \end{appendixproof}
\end{lemma}

\begin{lemma}\label{lem:Partial order on contexts}
  The relation on contexts \( D \sqsubseteq C \) defined as the existence of some context \( E \) such that \( C = D \circ E \) is a partial order.
  Furthermore, if two unrelated contexts \( C \) and \( D \) are equal for terms \( M \) and \( N \), i.e. \( C[M] = D[N] \), then \( M \unlhd D[X] \) for any term \( X \).
  \begin{appendixproof}
    \phantom{-}
    \paragraph{Reflexivity}
    Clearly, \( C \sqsubseteq C \) witnessed by the trivial context.

    \paragraph{Antisymmetry}
    Suppose \( C \sqsubseteq D \) and \( D \sqsubseteq C \).
    That is, \( C = D \circ E \) and \( D = C \circ F \).
    In particularly, we have that \( C = C \circ E \circ F \).
    It follows that both \( E \) and \( F \) are trivial.
    Thus, \( C = D \) as required.

    \paragraph{Transitivity}
    If \( C \sqsubseteq D \) and \( D \sqsubseteq E \), we have that \( C = D \circ F \) and \( D = E \circ G \) thus \( C = E \circ G \circ F \).
    Therefore, \( C \sqsubseteq E \) as required.

    \paragraph{Unrelated Context}
    Let \( C \) and \( D \) be two context unrelated by the partial order but such that \( C[M] = D[N] \) for any two terms \( M \) and \( N \).
    In particular, neither context can be trivial.
    We shall show by induction on the structure of the term \( C[M]\ (= D[N]) \), that \( M \) is a subterm of \( D[X] \) for all terms \( X \).
    \begin{itemize}
      \item Suppose it is a variable, constructor, or function symbol. Then both \( C \) and \( D \) must be trivial and thus related. This case, therefore, holds vacuously.
      \item Suppose it is an application \( L\ R \). Then there four subcases to consider:
            \begin{itemize}
              \item If \( M = L\ R \) or \( N = L\ R \), then either \( C \) or \( D \) are trivial and so this case is also holds vacuously.
              \item If \( C[\cdot] = L\ C'[\cdot] \) and \( D[\cdot] = L\ D'[\cdot] \), then we have two contexts that are also unrelated. Otherwise, if \( C' = D' \circ E \) for some \( E \), we'd have that \( C = D \circ E \) (and similarly if \( D' = C' \circ E \)). Furthermore, \( C'[M] = D'[N] \). And, by induction, \( M \) is a subterm of \( D'[X] \) for all terms \( X \). It therefore follows that \( X \) is a subterm of \( D[M] \) as required.
              \item When \( C[\cdot] = C'[\cdot]\ R \) and \( D[\cdot] = D'[\cdot]\ R \) the proof is analogous.
              \item If \( C[\cdot] = L\ C'[\cdot] \) and \( D[\cdot] = D'[\cdot]\ R \), then we have that \( C[M] = L\ C'[M] = D'[N]\ R = D[N] \) and, in particular, that \( L = D'[N] \) and \( R = C'[M] \). As \( R \) is clearly a subterm of \( D[X] \) for all terms \( X \), we are done.
              \item When \( C[\cdot] = C'[\cdot]\ R \) and \( D[\cdot] = L\ D'[\cdot] \) the proof is analogous.
            \end{itemize}
    \end{itemize}
  \end{appendixproof}
\end{lemma}

\defn{Substitutions}, typically \( \theta \), are partial functions from variables to terms with the usual action $M\theta$ on terms $M$.
We write \( \theta_1 \circ \theta_0 \) for the composition of substitutions, defined as \( x \mapsto (\theta_0(x))\theta_1 \).

A \defn{stable order} on terms \( \leq \) is a partial order such that \( M\theta \leq N\theta \) follows from \( M \leq N \) for any substitution \( \theta \).

A \defn{rewrite rule} is a pair of terms, written \( M \rightarrow N \) such that \( M \) is of the form \( f\ M_0\ \cdots\ M_n \) where \( f \in \Sigma_\textrm{def} \), each \( M_i \) doesn't contain any defined function symbols, and both \( \Gamma \vdash M : d \) and \( \Gamma \vdash N : d \) for some type environment \( \Gamma \) and a datatype \( d \).

For a set of rewrite rules \( R \), we define the \defn{one-step reduction} as \( C[M\theta] \rightarrow_R C[N\theta] \) whenever \( M \rightarrow N \in R \).
We write \( M \rightarrow_R^* N \) for the reflexive-transitive closure of this relation.

A term \( M \) is in \( R \)-\defn{normal form} when there does not exist a term \( N \) such that \( M \rightarrow_R N \).
We write \( \norm{M} \) for the term \( N \) that is a normal form such that \( M \rightarrow_R^* N \).

\begin{remark}[Assumptions]\label{rem:Assumptions}
  We will assume some fixed set of rules \( R \) such that the induced relation \( {\rightarrow_R^*} \) is:
  \begin{itemize}[leftmargin=*]
    \item \emph{Complete}, in the sense that, no closed first-order term headed by a defined function symbol is in normal form. That is, for any term \( \emptyset \vdash f\ M_0\ \cdots\ M_n : d \) with \( f \in \Sigma_\textrm{def} \), there exists some \( N \) such that \( f\ M_0\ \cdots\ M_n \rightarrow_R N \).
    \item And both weakly normalising and confluent so that \( \norm{[\cdot]} \) is a well-defined function on terms.
  \end{itemize}
\end{remark}

It is easy to ensure that the rewriting system corresponding to a functional program is complete and is often guaranteed by compilers.
Pure functional programs are also confluent.
On the other hand, the assumption that the program is weakly normalising is not without loss of generality.
However, it has been observed that problems of non-termination are relatively rare in comparison to those of functional correctness. It is also worth noting that although undecidable, practical algorithms exist for verifying this property.

\begin{example}
  The reduction relation \( \rightarrow_R \) induced by the following program clearly satisfies the assumptions of \cref{rem:Assumptions}.
  \begin{lstlisting}
  add Zero     y = y
  add (Succ x) y = Succ (add x y)

  map f Nil         = Nil
  map f (Cons x xs) = Cons (f x) (map f xs)
  \end{lstlisting}
\end{example}

An \defn{equation}, generically \( \phi \) or \( \psi \), is an \emph{unordered} pair of terms \( M \) and \( N \) such that \( \Gamma \vdash M,\, N : d \) for a type environment \( \Gamma \) and datatype \( d \).
Equations are written as \( \Gamma \vdash M \doteq N \), or equivalently \( \Gamma \vdash N \doteq M \).
When clear from the context, we will omit the type enironment.

A \defn{(ground) instance} of an equation \( \Gamma \vdash M \doteq N \) is a subsitution \( \alpha \), such that, \( \emptyset \vdash \alpha(x) : \tau \) for all \( x : \tau \in \Gamma \).
An equation \( \Gamma \vdash M \doteq N \) is \defn{satisfied} by an instance, written \( \alpha \vDash M \doteq N \), if \( \norm{M\alpha} = \norm{N\alpha} \).
If it is satisfied by all such instances, then we say it is \defn{valid} and write \( \vDash M \doteq N \).

Note that the satisfaction relation, \( \cdot \vDash \cdot \), and by extension validity, is well-defined as normalisation is a function and syntactic equality is a symmetric relation.

\section{Cyclic Proofs}\label{sec:cyclic-proofs}

\begin{figure*}[ht]
  \centering
  \begin{minipage}{1\textwidth}
    \[
      \arraycolsep=11pt
      \begin{array}{cc}
        \prftree[l]
        {\( \rlnm{Refl} \)}
        {\Gamma \vdash M \doteq M}

         &
        \prftree[l, r]
        { \( (M \rightarrow_R^* M',\, N \rightarrow_R^* N') \) }
        {\( \rlnm{Reduce} \)}
        {\Gamma \vdash M' \doteq N'}
        {\Gamma \vdash M \doteq N}
        \\[20pt]

        \prftree[l]
        {\( \rlnm{Subst} \)}
        {\Delta \vdash M \doteq N}
        {\Gamma \vdash C[N\theta] \doteq P}
        {\Gamma \vdash C[M\theta] \doteq P}
        
         &
        \prftree[l]
        {\( \rlnm{Case} \)}
        {\forall k \in \Sigma_\textrm{cons}(d)}
        {\Gamma, \Delta \vdash M[k\ x_0\ \cdots\ x_n/x] \doteq N[k\ x_0\ \cdots\ x_n/x]}
        {\Gamma,\, x : d \vdash M \doteq N}
      \end{array}
    \]
  \end{minipage}
  \Description{}
  \caption{The inference rules for preproofs}\label{fig:Inference rules}
\end{figure*}

An infinitary proof generalises traditional finite derivation trees to possibly infinite ones.
Such proofs are not necessarily sound; the standard approach is, therefore, to first define \emph{preproofs}, which are later refined by a global condition to ensure the argument is well-founded~\cite{brotherston2006sequent}.

Cyclic proofs are a subclass of infinitary proofs whose derivation trees are regular, i.e.\ there are only finitely many distinct subtrees.
Such proofs can be represented as finite but incomplete derivation trees where unjustified premises called ``buds'' refer to other vertices called ``companions''~\cite{brotherston2005cyclic}.
We will, however, present the cycles of a preproof directly.

\begin{definition}\label{def:Preproof}
  A \defn{(cyclic) preproof} is a tuple \( P = (V,\, e,\, r,\, p) \) where \( V \) is a finite set of vertices, typically an initial segment of the natural numbers, such that, for each vertex \( v \in V \):
  \begin{itemize}[leftmargin=*]
    \item There is an associated equation \( e(v) \), inference rule from \cref{fig:Inference rules} \( r(v) \), and a finite sequence of vertices \( p(v) \in V^* \) called the premises. We write \( p_i(v) \) for the \( i^\textrm{th} \) element of \( p(v) \) starting with \( p_0 \)
    \item And
          \[
            \prftree
            {e(p_0(v))}
            {\dots}
            {e(p_n(v))}
            {e(v)}
          \]
          is a well-formed instance of the rule \( r(v) \).
  \end{itemize}

  The \defn{underlying graph} of a preproof \( P = (V,\, e,\, r,\, p) \) is a graph \( G(P) = (V,\, E) \), over the same set of vertices, where:
  \[
    E \defeq \{ (v,\, p_i(v)) \mid v \in V,\, p_i(v) \textrm{ is defined} \}
  \]

  Note that when a premise appears as part of a cycle, it needn't be a direct ancestor.
  In particular, a cousin node may be used as a lemma by the \( \rlnm{Subst} \) rule.
\end{definition}

\begin{remark}[Rules in \cref{fig:Inference rules} defining equational preproofs]
  \phantom{-}
  \begin{enumerate}
    \item The rules are named according to their goal-orientated use. Hence \( \rlnm{Reduce} \) refers to the reduction of a goal to the premise.
    \item In this light, we will refer to the left- and right-hand premises the \( \rlnm{Subst} \) rule as the \defn{lemma} and \defn{continuation} respectively.
          The reason behind this convention will later become apparent when discussing our proof search algorithm \cref{sec:Implementation}.
    \item As the usual rules of transitivity, congruence, and instantiation are instances of \( \rlnm{Subst} \), they are trivially derivable. Symmetry follows immediately from the use of unordered equations. In particular, any combination of these rules can be used to form cycles.
    \item In the rule \( \rlnm{Case} \), there are as many equations in the premises as there are constructors of the datatype \( d \).
  \end{enumerate}
\end{remark}

A trivial example of a preproof can be constructed by using substitution to rewrite any equation according to itself, thus assuming the exact equation which is to be proved.

\begin{example}\label{ex:Unsound preproof}
  Let \( V = \{ 0,\, 1 \} \), let \( e(0) \) and \( e(1) \) be the equations \( \mathtt{Cons}\ x\ xs \doteq \mathtt{Nil} \) and \( \mathtt{Nil} \doteq \mathtt{Nil} \), let \( r(0) \) and \( r(1) \) be the rules \( \rlnm{Subst} \) and \( \rlnm{Refl} \), and let \( p(0) = [0,\, 1] \) and \( p(1) = [] \).
  Then \( (V,\, e,\, r,\, p) \) is a preproof satisfying \cref{def:Preproof}. 

\end{example}

\begin{remark}[Representing preproofs]
  \phantom{-}
  \begin{enumerate}
    \item
          Here, and in what follows, we will depict preproofs as a set of finite trees with labelled vertices and ``back edges'' that reference those labels.
          For example, the preproof of \cref{ex:Unsound preproof} would be presented as follows:
          \[
            \prftree[l]
            {\( \rlnm{Subst} \)}
            {
              \textrm{(0)}
            }
            {
              \prftree[l]
              {\( \rlnm{Refl} \)}
              {\mathtt{Nil} \doteq \mathtt{Nil}}
            }
            {\textrm{0: } \mathtt{Cons}\ x\ xs \doteq \mathtt{Nil}}
          \]
    \item To keep proofs compact, we shall also omit vertices justified by \( \rlnm{Reduce} \).
  \end{enumerate}
\end{remark}

Although this example clearly illustrates that preproofs are not necessarily sound arguments, they are, however, \emph{locally} sound in the sense that the premises of an inference rule 
justify its conclusion.
This property is witnessed by relating instances of a vertex to those of its premises, which is sufficient for concluding that the vertex's equation is satisfied for that instance. 

\begin{definition}
  Let \( (V,\, e,\, r,\, p) \) be a cyclic preproof with vertex \( v \in V \).
  Then for any instance of \( e(v) \), \( \alpha \), a \defn{preceding instance} is a pair \( (i,\, \beta) \), where \( p_i(v) \) is a premise and \( \beta \) is an instance of \( e(p_i(v)) \) such that one of the following conditions is met depending on the rule \( r(v) \):
  \begin{itemize}
    \item \( \rlnm{Case} \) where \( x : d \) is the variable upon which case analysis is performed. In this case, if \( \norm{(\alpha(x))} \) is of the form \( k\ M_0\ \cdots\ M_n \) and \( p_i(v) \) is the premise associated with the constructor \( k \) using fresh variables \( x_0,\, \dots,\, x_n \), then \( (i,\, \beta) \) is a preceding instance where \( \beta \) is defined as follows:
          \[
            \beta(y) \defeq \begin{cases}
              M_i       & y = x_i  \\
              \alpha(y) & y \neq x \\
            \end{cases}
          \]
    \item \( \rlnm{Subst} \) with substitution \( \theta \). In this case, \( (0, \, \alpha \circ \theta) \) and \( (1,\, \alpha) \) are preceding instances for the lemma and continuation respectively.
    \item Otherwise, there is a unique premise, for which \( (0,\, \alpha) \) is a preceding instance.
  \end{itemize}
\end{definition}

The following lemma states the important property that preceding instances must witness --- the contrapositive of local soundness (i.e.\ from an invalid conclusion, one can derive an invalid premise).

\begin{lemma}\label{lem:Unsatisfied preceding instance}
  Let \( (V,\, e,\, r,\, p) \) be a cyclic preproof with vertex \( v \in V \).
  If \( \alpha \) is an instance of \( e(v) \) such that \( \alpha \not\vDash e(v) \), then there exist a preceding instance \( (i,\, \beta) \) where \( \beta \not\vDash e(p_i(v)) \).
  \begin{appendixproof}
    Consider the possible inference rules \( r(v) \):
    \begin{itemize}
      \item \( \rlnm{Refl} \) Suppose the conclusion \( \Gamma \vdash M \doteq M \) is not satisfied by some ground instance \( \alpha \). Then \( \norm{M\alpha} \neq \norm{M\alpha} \), which immediately gives us a contradiction.
      \item \( \rlnm{Reduce} \) Suppose \( \alpha \not\vDash M \doteq N \), i.e. \( \norm{M\alpha} \neq \norm{N\alpha} \). We have that, \( M\alpha \rightarrow_R^* M'\alpha \) and \( N\alpha \rightarrow_R^* N'\alpha \). And, therefore, \( \norm{M'\alpha} \neq \norm{N'\alpha} \) by cofluence. Hence \( \alpha \not\vDash M' \doteq N' \). As \( (0,\, \alpha) \) is a preceding instance we are done.
      \item \( \rlnm{Subst} \) Suppose the lemma \( \vdash M \doteq N \) is satisfied by \( \beta = \alpha \circ \theta \), i.e. \( \norm{M\beta} = \norm{N\beta} \), but the conclusion \( \vdash C[M\theta] \doteq P \) is not satisfied by \( \alpha \). We have that \( \norm{M\beta} = \norm{N\beta} \) and, therefore, that  \( \norm{C[M\theta]\alpha} = \norm{C[N\theta]\alpha} \). It follows that \( \alpha \not\vDash C[N\theta] \doteq P \) and hence the preceding instance \( (\alpha,\, 1) \) meets the requirements. On the other hand, if the lemma \( \vdash M \doteq N \) is not satisfied by \( \beta \), then we are immediately done as this is also a preceding instance.
      \item \( \rlnm{Case} \) Suppose \( \vdash M \doteq N \) is not satisfied by \( \alpha \). As \( R \) is complete, the term assigned to \( x \) under \( \alpha \) must normalise to a term of the form \( k\ M_0\ \cdots\ M_n \) for some \( k \). Let \( p_i(v) \) be the premise associated the constructor \( k \). We claim that the preceding instance \( (\beta,\, i) \), where
            \[
              \beta(y) \defeq \begin{cases}
                M_i       & y = x_i  \\
                \alpha(y) & y \neq x \\
              \end{cases},
            \]
            meets the requirements. As \( \norm{M[k\ x_0\ \cdots\ x_n/x]\alpha_1} = \norm{M\alpha_0} \), which is not equal to \( \norm{N[k\ x_0\ \cdots\ x_n/x]\alpha_1} = \norm{N\alpha_0} \), we have that \( \beta \not\vDash e(p_i(v)) \) as required.
    \end{itemize}
  \end{appendixproof}
\end{lemma}

There are two direct corollaries of this lemma.
The first is that the equation of each vertex in a cyclic preproof is valid if all of its premises are:

\begin{corollary}[Local soundness]\label{lem:Local soundness}
  Let \( (V,\, e,\, r,\, p) \) be a cyclic preproof with vertex \( v \in V \).
  If the equation of each premise is valid, i.e. \( \vDash e(p_i(v)) \) when \( p_i(v) \) is defined, then \( \vDash e(v) \).
  \begin{appendixproof}
    Suppose, for the purpose of contradiction, that \( e(v) \) is invalid.
    Let \( \alpha \) be an instance that it does not satisfy.
    Then, by the \cref{lem:Unsatisfied preceding instance}, there is a preceding instance \( (i,\, \beta) \) such that \( \beta \not\vDash e(p_i(v)) \).
    However, as this equation is assumed to be valid, we have a contradiction.
    It follows that \( e(v) \) is valid.
  \end{appendixproof}
\end{corollary}

However, \cref{lem:Unsatisfied preceding instance} also implies that we can extract an infinite sequence of invalid equations from any invalid equation in a cyclic preproof.
This process also gives us the corresponding instances that are not satisfied.

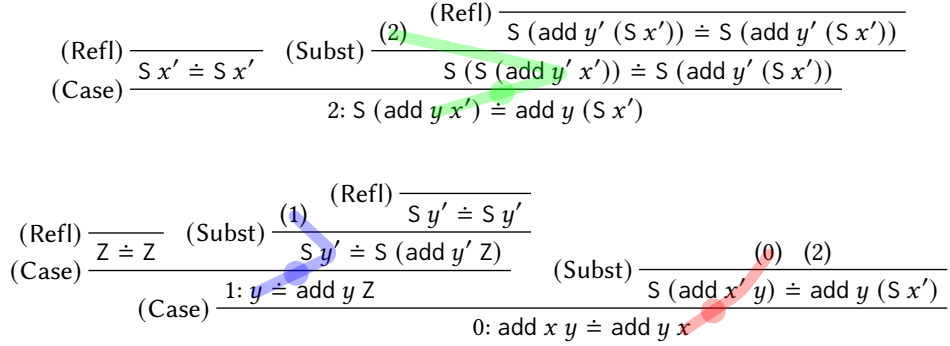
\begin{figure*}[ht]
  \begin{minipage}{1\textwidth}
    \[
      \begin{array}{c}
        \prftree[l]
        {\( (\textsf{Case}) \)}
        {
          \prftree[l]
          {\( (\textsf{Refl}) \)}
          {
            \mathtt{S}\ x' \doteq \mathtt{S}\ x'
          }
        }
        {
          \prftree[l]
          {\( (\textsf{Subst}) \)}
          {
            \textrm{(2\tikzmark{y5})}
          }
          {
            \prftree[l]
            {\( (\textsf{Refl}) \)}
            {\mathtt{S}\ (\mathtt{add}\ y'\ (\mathtt{S}\ x')) \doteq \mathtt{S}\ (\mathtt{add}\ y'\ (\mathtt{S}\ x'))}
          }
          {
            \mathtt{S}\ (\mathtt{S}\ (\mathtt{add}\ y'\tikzmark{y4}\ x')) \doteq \mathtt{S}\ (\mathtt{add}\ y'\ (\mathtt{S}\ x'))
          }
        }
        {
          \textrm{2: } \mathtt{S}\ (\mathtt{add}\ y\tikzmark{y3}\ x') \doteq \mathtt{add}\ y\ (\mathtt{S}\ x')
        }
        \\[20pt]
        \prftree[l]
        {\( (\textsf{Case}) \)}
        {
          \prftree[l]
          {\( (\textsf{Case}) \)}
          {
            \prftree[l]
            {\( (\textsf{Refl}) \)}
            {\mathtt{Z} \doteq \mathtt{Z}}
          }
          {
            \prftree[l]
            {\( (\textsf{Subst}) \)}
            {
              \textrm{(1\tikzmark{y2})}
            }
            {
              \prftree[l]
              {\( (\textsf{Refl}) \)}
              {\mathtt{S}\ y' \doteq  \mathtt{S}\ y'}
            }
            {\mathtt{S}\ y'\tikzmark{y1} \doteq  \mathtt{S}\ (\mathtt{add}\ y'\ \mathtt{Z})}
          }
          {
            \textrm{1: } y\tikzmark{y0} \doteq \mathtt{add}\ y\ \mathtt{Z}
          }
        }
        {
          \prftree[l]
          {\( (\textsf{Subst}) \)}
          {
            \textrm{(0\tikzmark{x2})}
          }
          {
            \textrm{(2)}
          }
          {
            \mathtt{S}\ (\mathtt{add}\ x'\tikzmark{x1}\ y) \doteq \mathtt{add}\ y\ (\mathtt{S}\ x')
          }
        }
        {\textrm{0: } \mathtt{add}\ x\ y \doteq \mathtt{add}\ y\ x\tikzmark{x0}}
      \end{array}
    \]
  \end{minipage}

  \begin{tikzpicture}[overlay, remember picture]
    \draw[trace, red]{(pic cs:x0) -- node[progress] {} (pic cs:x1) -- (pic cs:x2)};
    \draw[trace, green] {(pic cs:y3) -- node[progress] {} (pic cs:y4) -- (pic cs:y5)};
    \draw[trace, blue] {(pic cs:y0) -- node[progress] {} (pic cs:y1) -- (pic cs:y2)};
  \end{tikzpicture}

  \Description{}
  \caption{A cyclic proof that addition is commutative.}\label{fig:Commutative addition}
\end{figure*}

If a parallel sequence of terms can be constructed from these instances that is infinitely decreasing according to some well-founded order, we have shown there are no invalid equations.
To this end, a global condition is placed upon cyclic preproofs based on the notion of a \emph{trace}.
A trace is another sequence of terms intuitively capturing any dependency between the instances of a conclusion and its premise.

\begin{definition}
  A \defn{path} through a preproof \( P = ( V,\, e,\, r,\, p) \) is a finite or infinite sequence of vertices \( (v_i) \) such that, for each \( i \in \mathbb{N} \), \( v_{i+1} \) is a premise of \( v_i \), i.e.\ there is some \( j \) such that \( v_{i+1} = p_j(v_i) \).
\end{definition}

\begin{definition}
  Let \( {\leq} \) be a stable, well-founded order.
  A \( {\leq} \)-\defn{trace} along a path \( (v_i) \) is a finite or infinite sequence of terms \( (T_i) \) where \( T_{i+1} \) is constrained according to the rule \( r(v_i) \):
  \begin{itemize}
    \item \( \rlnm{Case} \) where \( x : d \) is the variable upon which case analysis is performed. If \( v_{i+1} \) is the premise associated with constructor \( k \) using fresh variables \( x_0,\, \dots,\, x_n \), then \( T_{i+1} \leq T_i[k\ x_0\ \cdots\ x_n/x] \).
    \item \( \rlnm{Subst} \) with substitution \( \theta \). If \( v_{i+1} \) is the lemma, then \( T_{i+1}\theta \leq T_i \) and if \( v_{i+1} \) is the continuation, then \( T_{i+1} \leq T_i \).
    \item Otherwise, \( T_{i+1} \leq T_i \)
  \end{itemize}

  When there is a strict inequality in the above definition, we say that \( v_i \) is a \defn{progress point}.
\end{definition}

\begin{remark}[Progress points]
  Part of our intention with this work is to relate rewriting induction to cyclic proofs.
  It is, therefore, not possible to build a specific relationship between derivations and progress points, as is done in e.g. Brotherston's work~\cite{brotherston2006sequent}, because different rules will entail a progress point for different orderings.
  For example, our implementation is based on the substructural order where progress points are marked by the \( \rlnm{Case} \) rule, but a reduction order would also use \( \rlnm{Reduce} \) and \( \rlnm{Subst} \) as in \cref{sec:Rewriting induction}.
\end{remark}

\begin{lemma}\label{lem:Monotonic traces}
  Let \( (V,\, e,\, r,\, p) \) be a preproof with vertex \( v \in V \), let \( \alpha \) be an instance of \( e(v) \), and let \( (i,\, \beta) \) be a preceding instance.
  If \( T_0,\, T_1 \) is a trace for the path \( v,\, p_i(v) \), then \( T_1\beta \leq T_0\alpha \) and, in particular, \( T_1\beta < T_0\alpha \), if \( v \) is a progress point.
  \begin{appendixproof}
    Consider the possible cases that define a preceding instance:
    \begin{itemize}
      \item If \( r(v) \) is \( \rlnm{Reduce} \), then \( \beta \) is \( \alpha \) and \( T_1 \leq T_0 \). Thus by stability \( T_1\beta \leq T_0\alpha \) as required.
      \item If \( r(v) \) is \( \rlnm{Subst} \) and \( i = 1 \), i.e. \( p_i(v) \) is the continuation, then the same argument applies.
      \item If \( r(v) \) is \( \rlnm{Subst} \) and \( i = 0 \), i.e. \( p_i(v) \) is the lemma, then \( \beta = \alpha \circ \theta \). From the definition of a trace, we have that \( T_1\theta \leq T_0 \). Thus \( T_1\theta\alpha \leq T_0\beta \) by stability, which is equivalent to \( T_1\alpha \leq T_0\alpha \) as required.
      \item Finally, if \( r(v) \) is \( \rlnm{Case} \), then \( \beta \) maps the fresh variables \( x_i \) to the arguments of \( \norm{(\alpha(x))} = k\ M_0\ \cdots\ M_n \). From the definition of a trace, we have that \( T_1 \leq T_0[k\ x_0\ \cdots\ x_n/x] \). And so by satibility, \( T_1\beta \leq T_0\alpha \), as required.
    \end{itemize}
    Clearly, each of the preceding cases generalises to strict inequality.
  \end{appendixproof}
\end{lemma}

The aforementioned lemma shows how a trace, as previously defined, leads to a monotonic sequence of terms by closing those terms according to a sequence of preceding instances that emerges as a consequence of  \cref{lem:Unsatisfied preceding instance}.
If every path has a trace, then we can construct an infinitely decreasing sequence of terms for any sequence of invalid equations generated by \cref{lem:Unsatisfied preceding instance}.
Thus we define a proof as a preproof that satisfies the following \emph{global correctness condition}.

\begin{definition}\label{def:Proof}
  A \({\leq}\)-\defn{(cyclic) proof} is a preproof such that, for every infinite path \( (v_i) \), there is a suffix, i.e. \( (v_{i+k}) \) for some \( k \in \mathbb{N} \), which has a \({\leq}\)-trace with infinitely many progress points.
\end{definition} 

Because our definition of a trace and its progress points is highly generic, it is undecidable if a sufficient set of traces exists or not.
This is a significant difference from proofs by structural induction, whose validity is effectively a syntactic well-formedness condition.
Although undesirable, we will, in practice, restrict the space of traces according to a particular application as in \cref{sec:Algorithmic system}.
In particular, our implementation only checks for traces composed solely of variables, which is decidable.
However, we chose not to overfit the declarative system as alternative restrictions are equally valid.
This point is discussed further under related work.

\begin{example}[Commutativity of addition]
  \cref{fig:Commutative addition} displays a preproof for the commutativity of addition.
  Note that there are implicit applications of the \( \rlnm{Reduce} \) rule in the \( \mathtt{S} \)-case of the root node, i.e.\ the rewriting of \( \mathtt{add}\ (\mathtt{S}\ x)\ y \) to \( \mathtt{S}\ (\mathtt{add}\ x'\ y) \) in it's left parent.
  And similarly throughout.

  To show that this is also a \( {\unlhd} \)-proof, we must consider every infinite path and show they each have a suffix with an infinitely progressing trace.
  There are three cycles we must consider:
  \begin{itemize}
    \item One passing through \( 0 \) by following the continuation in case associated with the \( \mathtt{S} \) constructor, for which the trace \( x,\, x',\, x,\, x,\, \dots \) is sufficient. The decrease \( x' \lhd x[S\ x'/x] \) marks a progress point.
    \item One passing through \( 1 \) by following the lemma in the case associated with the \( \mathtt{Z} \) constructor, for which the trace \( y,\, y',\, y,\, y,\, \dots \) is similarly sufficient.
    \item And finally, one passing through \( 2 \) by following the lemma in case associated with the \( \mathtt{S} \) constructor, for which the trace \( y,\, y',\, y,\, y,\, \dots \) is also sufficient.
  \end{itemize}
  These traces are informally depicted as coloured lines in the preproof diagram with progress points marked by circles, following~\cite{kuperberg2021cyclic}.
\end{example}

\begin{theorem}[Global soundness]\label{thm:Global soundness}
  Let \( {\leq} \) be a stable well-founded order.
  If \( (V,\, e,\, r,\, p) \) is a \({\leq}\)-proof with some vertex \( v \in V \), then \( \vDash e(v) \).
  \begin{appendixproof}
    Suppose, for the purpose of contradiction, that \( e(v) \) is invalid.
    Let \( \alpha \) be an instance that it does not satisfy.
    We construct a path and an infinite sequence of instances by induction:
    \begin{itemize}
      \item Let \( v_0 = v \) and \( \alpha_0 = \alpha \).
      \item Suppose \( v_i \) and \( \alpha_i \) are defined such that \( \alpha_i \not\vDash e(v_i) \). Then, by \cref{lem:Unsatisfied preceding instance}, there exists a preceding instance \( (j,\, \beta) \) and let \( v_{i+1} = p_j(v_i) \) and \( \alpha_{i+1} = \beta \).
    \end{itemize}

    There exists some \( k \in \mathbb{N} \) and a trace \( (T_i) \) for the path \( (v_{i+k}) \) by \cref{def:Proof}.
    Consider the ground instances \( (T_i\alpha_{i+k}) \).
    It follows from \cref{lem:Monotonic traces} that this sequence is monotonic with respect to \( {\leq} \).
    Furthermore, there are infinitely many progress points, i.e. \( i \) such that \( T_{i+1}\alpha_{i+ k +1} < T_i\alpha_{i + k} \).
    If we consider the subsequence of progress points, then we have a strictly decreasing chain.
    By assumption the that \( {\leq} \) is well-foundned, this gives us a contradiction.
    Thus \( \vDash e(v) \) as required.
  \end{appendixproof}
\end{theorem}

\begin{remark}[A refinement of global correctness]
  Allowing for traces that only cover a certain suffix of a path is particularly useful in the context of cyclic proofs, as paths must be ultimately periodic.
  It is, therefore, only necessary to find a trace for every cycle.
  Note, however, cycles may overlap, and so it is \emph{not} sufficient to assign a single term to each vertex.
\end{remark}

\section{Rewriting Induction}\label{sec:Rewriting induction}

\begin{figure*}[ht]
  \begin{minipage}{1\textwidth}
    \[
      \begin{array}{cc}
        \prftree[l]
        { \( (\textsf{End}) \) }
        { \vdash (\emptyset,\, H) }

         &
        \prftree[l]
        { \( (\textsf{Delete}) \) }
        { \vdash (E,\, H) }
        { \vdash (E \cup \{ M = M \},\, H) }
        \\[20pt]

        \prftree[l, r]
        { \( (M \rightarrow_{R \cup H}^* M') \) }
        { \( (\textsf{Simplify}) \) }
        { \vdash (E \cup \{ M' = N \},\, H) }
        { \vdash (E \cup \{ M = N \},\, H) }

         &
        \prftree[l, r]
        { \( (N < M) \) }
        { \( (\textsf{Expand}) \)}
        { \vdash (E \cup \textsf{Expand}_C(M = N),\, H \cup \{ M \rightarrow N \}) }
        { \vdash (E \cup \{  M = N \},\, H) }
      \end{array}
    \]
  \end{minipage}
  \Description{}
  \caption{Inference rules of rewriting induction.}\label{fig:Rewriting induction}
\end{figure*}

As discussed in the introduction, automating traditional inductive proofs is highly non-trivial, and many alternatives have thus been proposed.
One such well-developed line of work is proof by consistency or inductionless induction~\cite{musser1980proving, huet1982proofs, kapur1987proof}.
Musser observed that if an equation can be consistently added to a strongly complete theory, it is true of its least model.
The consistency of an equational theory, in this case that \( \not\vdash \mathtt{False} = \mathtt{True} \), can then be verified by converting the theory into a confluent and terminating rewrite system by using the Knuth-Bendix algorithm~\cite{knuth1983simple}.

Rewriting induction, due to Reddy, highlights its core mechanisms~\cite{reddy1990term}.
The principal idea behind rewriting induction is to perform induction using a well-founded ordering that includes the reduction relation --- a ``reduction'' order.
A reduction order is more flexible than the substructural order in that more terms are related.
Furthermore, unlike the use of a structural induction scheme, this approach can be easily extended to mutually inductive datatypes as there is no need to invent a complementary induction hypothesis.

For this section, we shall assume \( {\leq} \) is a \defn{reduction order}.
That is, a well-founded stable order such that each rewrite rule \( M \rightarrow N \in R \) is strictly decreasing, i.e. \( N < M \).

The \defn{decreasing order} \( {\prec} \) is defined as the transitive closure of the relation \( {<} \cup {\lhd} \).

\begin{lemma}
  \( {\prec} \) is a reduction order.
  \begin{appendixproof}
    \phantom{-}
    \paragraph{Well-foundedness}
    \( {\prec} \) is equivalent to \( {<} \cup {\lhd} \cup ({<} \circ {\lhd}) \) and, therefore, well-founded~\cite{reddy1990term}.

    \paragraph{Stability}
    Suppose \( M \prec N \).
    Therefore, there exists some \( n \in \mathbb{N} \) such that \( M \mathbin{{({<} \cup {\lhd})}^n} N \).
    We shall show that \( C[M\theta] \prec C[N\theta] \) by induction on \( n \).
    \begin{itemize}
      \item If \( M < N \), then clearly \( C[M\theta] \prec C[N\theta] \).
      \item If \( M < R \) and \( R \prec N \), then \( C[M\theta] < C[R\theta]\) by induction \( C[R\theta] \prec C[N\theta] \) and \( C[M\theta] \prec C[N\theta] \) as required.
      \item The \( {\lhd} \) case are analogous.
    \end{itemize}

    Thus \( {\prec} \) is well-founded, stable, and compatible with \( R \) as it subsumes \( < \),
  \end{appendixproof}
\end{lemma}

A serious complication of rewriting induction, however, is that all lemmas (including equations that play the role of induction hypotheses) must also be orientated according to the reduction order.
Properties such as the commutativity of addition are, therefore, difficult to prove.
Although there are extensions that allow for unoriented equations, the increase in complexity detracts from the advantage of rewriting induction --- its simplicity~\cite{aoto2006dealing}.

Furthermore, rewriting induction is highly sensitive to the choice of order and choosing an order in advance is a non-trivial task.
For example, if the term \( \mathtt{add}\ (\mathtt{add}\ x\ y)\ z \) is less than \( \mathtt{add}\ x\ (\mathtt{add}\ y\ z) \), then it is impossible to prove addition is associative without externally supplied lemmas.

Our cyclic proof system allows for both unoriented equations and is ambivalent to the choice of order, overcoming these limitations.
However, it is worth reiterating that we have not provided a method for verifying the global condition, which is required in the general case.

\begin{definition}
  The most significant inference rule concerns the \defn{expansion} of an equation:
  \[
    \begin{array}{l}
      \textsf{Expand}_C(C[f\ M_0\ \dots\ M_n] = N) \defeq \\
      \quad \{ C[L]\theta = N\theta \mid f\, N_0 \cdots N_n \rightarrow L \in R ,\, \theta = \textsf{mgu}(\overline{M},\, \overline{N}) \}
    \end{array}
  \]
  following the presentation used in~\cite{aoto2006dealing}.
\end{definition}

This operator is used to perform case analysis of the variables, which are instantiated with constructors.
However, a critical part of this definition is that a reduction step has occurred, and the left-hand side is, therefore, strictly smaller.
In other words, it marks a progress point for a reduction order.

\begin{definition}[Rewriting induction]
  The inference rules of rewriting induction manipulate pairs \( (E,\, H) \) of oriented equations \( E \) (denoted \( M = N \), in contrast to \( M \doteq N \)) to be proven, and rewrite rules \( H \) that supplement the original set \( R \).
  The judgement \( \vdash (E,\, H) \) is inductively defined by the rules of \cref{fig:Rewriting induction}.

  Note that although the rules from \( H \) must comply with the reduction order, they needn't be orthogonal to \( R \) or behave like a functional program.
  For example, \( \mathtt{add}\ (\mathtt{add}\ x\ y)\ z \rightarrow \mathtt{add}\ x\ (\mathtt{add}\ y\ z) \) is valid despite there already being rules that govern the reduction of \( \mathtt{add} \).
\end{definition}

\begin{theorem}[Soundness]
  If \( {} \vdash (E,\, \emptyset) \) is a rewriting induction derivation, then every equation in \( E \) is valid~\cite{reddy1990term}.
\end{theorem}

\subsection{Translation to Cyclic Proof}

Rewriting induction allows for previously seen equations to be used as hypotheses.
This circularity is not unsound as hypotheses are only introduced through a strict decrease.

We will show that rewriting induction proofs can be translated into our cyclic proof system and, therefore, can be seen as a form of cyclic proof search, see \cref{thm:Rewriting induction}.
Furthermore, as rewriting induction subsumes inductionless induction, a line of work that adapts the Knuth-Bendix completion procedure to perform saturation based proofs by consistency, our system also subsumes that approach~\cite{reddy1990term}.

We will construct a cyclic proof by induction over a rewriting induction derivation.
Cyclic proofs discharge their hypothesis globally rather than locally, and thus we need to allow for undischarged hypotheses when reasoning locally in this manner.
We first define this generalisation of cyclic proofs as follows:

\begin{definition}
  A \defn{partial proof} is a tuple \( (V,\, H,\, e,\, r,\, p) \) where \( V \) and \( H \) are disjoint finite sets of vertices such that:
  \begin{itemize}
    \item For each \( v \in V \cup H \), there is an associated equation \( e(v) \).
    \item For each \( v \in V \), there is inference rule from \cref{fig:Inference rules} \( r(v) \), and list of vertices \( p(v) \in {(V \cup H)}^* \) called the premises. We write \( p_i(v) \) for the \( i^\textrm{th} \) element of \( p(v) \) starting with \( p_0 \).
    \item And
          \[
            \prftree
            {e(p_0(v))}
            {\dots}
            {e(p_n(v))}
            {e(v)}
          \]
          is a well-formed instance of the rule \( r(v) \).
  \end{itemize}

  Furthermore, partial proofs must also satisfy the global condition that for every path \( (v_i) \), there is a suffix, i.e. \( (v_{i+k}) \) for some \( k \in \mathbb{N} \), which has a \({\preceq}\)-trace with infinitely many progress points.

  We will refer to the elements of \( H \) as \defn{hypotheses}.
\end{definition}

Intuitively, a partial proof is a proof where the hypotheses \( H \) may be used as premises but needn't be justified by an instance of an inference rule themselves.
Note that when \( H \) is empty, we have a cyclic proof.

\begin{theorem}\label{thm:Rewriting induction}
  If \( \vdash (E,\, H) \) is rewriting induction derivation, then there exists a partial proof \( (V',\, H',\, e,\, r,\, p) \) where \( E \subseteq \{ e(v) \mid v \in V \} \) and \( H = \{ e_1 \rightarrow e_2 \mid e_1 \doteq e_2 \in H',\, e_2 \preceq e_1 \}  \), i.e.\ the rewrite rules of \( H \) are orientations of equations in \( H' \).
  \begin{appendixproof}
    We shall proceed by induction over the deriviation of \( \vdash (E,\, H) \).

    \begin{itemize}
      \item \( (\textsf{End}) \) There exists a partial proof with no vertices and a set of hypotheses for each equation in \( H \).
      \item \( (\textsf{Delete}) \) By induction, we have a partial proof \( (V,\, H,\, e,\, r,\, p) \) that corresponds to a derivation of \( \vdash (E,\, H) \). Let \( v \not\in V \cup H \) be a fresh vertex. Then define \( e'(w) \) as \( M \doteq M \) when \( w = v \) and \( e'(w) = e(w) \) otherwise. Similarly, extend \( r \) and \( p \) such that \( r'(v) \) is \( \rlnm{Refl} \) and \( p'(v) = [] \). Then the preproof \( (\{ v \} \cup V,\, H,\, e',\, r',\, p') \) has the correct structure. Furthermore, there are no additional infinite paths, so we are done.
      \item \( (\textsf{Simplify}) \) By induction, we have a partial proof \( P = (V,\, H,\, e,\, r,\, p) \) with some vertex \( v \in V \) such that \( e(v) = M' \doteq N \) where \( M \rightarrow_{R \cup H}^* M' \). We shall now, by induction over the length of this reduction sequence, construct a partial proof \( P' = (V \cup V',\, H,\, e',\, r',\, p') \) with some vertex \( v' \in V \cup V' \) such that \( e'(v') = M \doteq N \).
            \begin{itemize}
              \item Suppose \( M = M' \), then we are immediate done with \( P' = (V \cup \emptyset,\, H,\, e,\, r,\, p) \).
              \item Suppose \( M \rightarrow_R M' \), and there is a partial proof \( P = (V,\, H,\, e,\, r,\, p) \) where \( e(v) = M' \doteq N \) for some \( v \in V \). Let \( v' \not\in V \cup H \) be a fresh vertex. Then defined \( e'(w) \) as \( M \doteq N \) when \( w = v' \) and \( e'(w) = e(w) \) otherwise. Similarly, extend \( r \) and \( p \) such that \( r'(v') \) is \( \rlnm{Reduce} \) and \( p'(v') = [v] \). Then the preproof \( P' = (V \cup (\{ v' \}),\, H,\, e',\, r',\, p') \) has the correct structure. Furthermore, any infinite path has a suffix that is an infinite path in the partial proof \( P \), and thus a trace with infinitely many progres points.
              \item Suppose \( M \rightarrow_H M' \), and there is a partial proof \( P = (V,\, H,\, e,\, r,\, p) \) where \( e(v) = M' \doteq N \) for some \( v \in V \). Let \( v' \not\in V \cup H \) be a fresh vertex. Then defined \( e'(w) \) as \( M \doteq N \) when \( w = v' \) and \( e'(w) = e(w) \) otherwise. Similarly, extend \( r \) and \( p \) such that \( r'(v') \) is \( \rlnm{Subst} \) and \( p'(v') = [h,\, v] \) where \( e(h) \in H \) is the hypotheses used. Then the preproof \( P' = (V \cup (\{ v' \}),\, H,\, e',\, r',\, p') \) has the correct structure. Furthermore, any infinite path has a suffix that is an infinite path in the partial proof \( P \), and thus a trace with infinitely many progres points.
            \end{itemize}
      \item \( (\textsf{Expand}) \) In this case, we have a partial proof \( P = (V,\, H \cup \{ h \},\, e,\, r,\, p) \) where \(e(h) = M \doteq N \) and, for each equation \( \phi \in \textsf{Expand}_C(M = N) \), there is some vertex \( v \in V \) such that \( e(v) = \phi \). From which, we must construct a partial proof \( P' = (V \cup \{ h \} \cup V',\, H,\, e',\, r',\, p') \) where \( e'(h) = M \doteq N \). Let \( E \subseteq V \) be a set of vertices that covers the equations \( \textsf{Expand}(M = N) \).

            As the head of rewrite rules are of the form \( f\ M_0\ \dots\ M_n \) where each \( M_i \) is built solely from constructors and the rules are complete, there exists a finite tree built from \( \rlnm{Case} \) and \( \rlnm{Reduce} \) from \( h \) to the vertices that cover the equations from \( \textsf{Expand}_C(M = N) \), let \( V' \) be those internal nodes and extend \( r' \) and \( p' \) accordingly. Furthermore, any path from \( v \) to a vertex from \( v' \in E \) can be assigned a trace \( T,\, \dots,\, T\theta \) if \( e(v') = C[L]\theta \doteq N\theta \) for some \( f\, N_0 \cdots N_n \rightarrow L \in R \) and \( \theta = \textsf{mgu}(\overline{M},\, \overline{N}) \).

            Suppose there is a infinite path in the partial proof \( P' \) that doesn't have a suffix in \( P \).
            Without loss of generality, it must start at \( h \), proceeds to a specific case \( v \in E \), and then cycle back to \( h \), which is used as a lemma in \( P \) by the \( \rlnm{Simplify} \) case.
            For this path, we consider the trace \( M,\, \dots,\, M\theta,\, L,\, \dots,\, M\), which has infinitely many progress points as \( M\theta \rightarrow_R C[L]\theta \) must be included in \( \prec \).
    \end{itemize}
  \end{appendixproof}
\end{theorem}

\begin{remark}
  Rewriting induction and related approaches do not typically require a confluent rewrite system.
  Therefore, this theorem only shows that our system subsumes rewriting induction when confluence is taken as an assumption, which is the case for our intended application.
  It is also worth noting that we only require confluence when defining the semantics of terms; we are confident that the proof system could be made sound for non-confluent rewrite systems.
\end{remark}

Cyclic proofs, for a generic sequent calculus, have been shown to subsume traditional structural induction~\cite{brotherston2005cyclic}.
Although this result is not directly applicable to our system, which is specialised to unconditional equational reasoning, we conjecture that an analogous argument could be made with unconstrained usage of the \( \rlnm{Subst} \) rule.
For examples of the translation from structural induction to proofs in our calculus, see the long version of this paper~\cite{jones2021cycleq}.

\begin{toappendix}
  \begin{example}
    \begin{figure*}[ht]
      \begin{minipage}{1\textwidth}
        \[
          \begin{array}{c}
            \prftree[l]
            {\( \rlnm{Ind} \)}
            {
              \prftree[l]
              {\( \rlnm{Refl} \)}
              {[\,] \doteq [\,]}
            }
            {
              \prftree[l]
              {\( \rlnm{Subst} \)}
              {
                \prftree[l]
                {\( \rlnm{Imp}\) }
                {\mathtt{map}\ \mathtt{id}\ xs \doteq xs \Rightarrow \mathtt{map}\ \mathtt{id}\ xs \doteq xs}
              }
              {
                \prftree[l]
                {\( \rlnm{Refl}\) }
                {\mathtt{map}\ \mathtt{id}\ xs \doteq xs \Rightarrow x : xs \doteq x : xs}
              }
              {\mathtt{map}\ \mathtt{id}\ xs \doteq xs \Rightarrow x : \mathtt{map}\ \mathtt{id}\ xs \doteq x : xs}
            }
            {\mathtt{map}\ \mathtt{id}\ xs \doteq xs}
          \end{array}
        \]
      \end{minipage}
      \Description{}
      \caption{A classical inductive proof that \( \mathtt{map}\ \mathtt{id}\ xs \doteq xs \).}\label{fig:Map induction}
    \end{figure*}

    \begin{figure*}[ht]
      \begin{minipage}{1\textwidth}
        \[
          \begin{array}{c}
            \prftree[l]
            {\( \rlnm{Case} \)}
            {
            \prftree[l]
            {\( \rlnm{Refl} \)}
            {\mathtt{map}\ \mathtt{id}\ [\,] \doteq \mathtt{map}\ \mathtt{id}\ [\,]}
            }
            {
            \prftree[l]
            {\( \rlnm{Subst} \)}
            {
              \textrm{(0\tikzmark{xs2})}
            }
            {
              \prftree[l]
              {\( \rlnm{Refl}\) }
              {x : xs' \doteq x : xs'}
            }
            {x : \mathtt{map}\ \mathtt{id}\ xs'\tikzmark{xs1} \doteq x : xs'}
            }
            {\textrm{0: } \mathtt{map}\ \mathtt{id}\ xs\tikzmark{xs0} \doteq xs}
          \end{array}
        \]
      \end{minipage}

      \begin{tikzpicture}[overlay, remember picture]
        \draw[trace, red]{(pic cs:xs0) -- node[progress] {} (pic cs:xs1) -- (pic cs:xs2)};
      \end{tikzpicture}

      \Description{}
      \caption{A cyclic proof that \( \mathtt{map}\ \mathtt{id}\ xs \doteq xs \).}\label{fig:Map cyclic}
    \end{figure*}

    \cref{fig:Map induction} shows a traditional inductive proof that \( \mathtt{map}\ \mathtt{id}\ xs \doteq xs \).
    This can be mechanically translated to a cyclic proof from our proof system, see \cref{fig:Map cyclic}, with the simple trace \( xs,\, xs',\, \dots \) that corresponds to doing induction over \( xs \).
  \end{example}

\end{toappendix}

\section{Detecting and Verifying Cycles}\label{sec:Algorithmic system}

Our proof system is designed to be used in a goal-orientated manner.
It is necessary to form cycles to produce a finite proof, and subsequently, verify that the global condition has been met.
In this section, we discuss these high-level aspects of our proof search algorithm.

There exists a generic cyclic theorem prover for sequent calculi --- the \textsc{Cyclist} system~\cite{brotherston2012generic}.
It is generic in that it supports an arbitrary set of inference rules.
Given this setup, it would be possible to na\"ively enumerate derivations of the goal formula and create cycles just when formulas are repeated.

Sequents discovered earlier in proof search are intuitively simpler in that they apply to a more general instance.
For example, consider the \( \rlnm{Case} \) rule where only certain instances of the conclusion will be relevant to each premise.
Therefore, it will often be necessary to generalise an equation to relate it to an ancestor.

However, it is desirable to avoid generalisation as part of normal proof search, as the space is often intractable and not guaranteed to lead to a cycle~\cite{maclean1999generalisation}.
The \textsc{Cyclist} framework is thus parameterised by a ``matching function'' that detects when cycles can be formed.
The matching function for first-order logic is a combination of weakening and substitution.
For separation logic, the matching function is the frame rule.

The capacity of the Cyclist framework for equational reasoning is known to be limited.
For example, the commutativity of addition cannot be automatically proven without the lemma \( \mathtt{add}\ x\ (\mathtt{S}\ y) \doteq \mathtt{S}\ (\mathtt{add}\ x\ y) \).
Our observation is that the existing matching functions are too restrictive for equational reasoning.

The usual rules for equational reasoning, i.e.\ the congruence axioms, are intractable due to the vast number of intermediate equations they create~\cite{bachmair1998equational}.
However, we cannot simply avoid such equational reasoning in cyclic proofs as they are often needed to form cycles.
We thus propose the substitution of equals as an alternative matching function, appearing in our proof system as the \( \rlnm{Subst} \) rule.
This way of closing cycles resembles the use of hypotheses as rewrite rules in rewriting induction.

Algorithmically, \( \rlnm{Subst} \) is only used as a matching function in a goal-directed manner.
The task of generating useful lemmas is a non-trivial and orthogonal concern~\cite{johansson2019lemma}.
For a given goal equation \( \Gamma \vdash M \doteq N \), any subterms that are instances of the left- or right-hand side of an existing node are considered and the goal is rewritten accordingly, leaving the continuation premise as a new subgoal.
Thus the lemma, i.e.\ the first premise of \( \rlnm{Subst} \), is always an equation that has already appeared in the tree, acting somewhat like an induction hypothesis.

There is a significant novelty in using \( \rlnm{Subst} \) as a matching function not present in the \textsc{Cyclist} system --- it doesn't completely close a branch of the derivation tree into a cycle but leaves a new subgoal, i.e.\ the continuation, which must also be solved.

\subsection{Refining Substitution}\label{sec:Refining substitution}

\begin{figure*}[ht]
  \begin{minipage}{1\textwidth}
    \[
      \scriptsize
      \begin{array}{ccc}\prftree[l]
        {\( \rlnm{Subst} \)}
        {
          \prftree[l]
          {\( \rlnm{Reduce} \)}
          {\Delta \vdash M \doteq N'}
          {\Delta \vdash M \doteq N}
        }
        {
          \prftree[l]
          {\( \rlnm{Reduce} \)}
          {\Gamma \vdash Q \doteq P'}
          {\Gamma \vdash C[N\theta] \doteq P}
        }
        {\Gamma \vdash C[M\theta] \doteq P}

         &
        \scalebox{1.5}{\( \rightsquigarrow \)}

         &
        \prftree[l]
        {\( \rlnm{Subst} \)}
        {
          \Delta \vdash M \doteq N'
        }
        {
          \prftree[l]
          {\( \rlnm{Reduce} \)}
          {\Gamma \vdash Q \doteq P'}
          {\Gamma \vdash C[N'\theta] \doteq P}
        }
        {\Gamma \vdash C[M\theta] \doteq P}
        \\[20pt]

        \prftree[l]
        {\( \rlnm{Subst} \)}
        {
          \prftree[l]
          {\( \rlnm{Subst} \)}
          {\Lambda \vdash M \doteq N}
          {\Delta \vdash D[N\theta] \doteq P'}
          {\Delta \vdash D[M\theta] \doteq P'}
        }
        {\Gamma \vdash C[P'\sigma] \doteq P}
        {\Gamma \vdash C[(D[M\theta])\sigma] \doteq P}

         &
        \scalebox{1.5}{\( \rightsquigarrow \)}

         &
        \prftree[l]
        {\( \rlnm{Subst} \)}
        {\Lambda \vdash M \doteq N}
        {
          \prftree[l]
          {\( \rlnm{Subst} \)}
          {\Delta \vdash D[N\theta] \doteq P'}
          {\Gamma \vdash C[P'\sigma] \doteq P}
          {\Gamma \vdash C[(D[N\theta])\sigma] \doteq P}
        }
        {\Gamma \vdash C[(D[M\theta])\sigma] \doteq P}
      \end{array}
    \]
  \end{minipage}
  \Description{}
  \caption{Redundancy of unreduced lemmas \& Reassociation of nested substitution.}\label{fig:Rewriting substitution}
\end{figure*}

While substitution is an appropriate technique for detecting cycles, it can also create many redundancies when searching for proofs that lead to performance issues.
Therefore, we only consider a subset of available lemmas for substitution in our implementation.
These are determined by the rule used to justify the lemma:
\begin{itemize}[leftmargin=*]
  \item \( \rlnm{Refl} \) Clearly, no useful lemma is justified by reflexivity as the continuation is identical to the goal.
  \item \( \rlnm{Reduce} \) We also do not consider lemmas justified by reduction. This restriction follows naturally from the reasonable strategy that we ought to reduce a goal as far as possible before further reasoning. Suppose we have a goal \( C[M\theta] \doteq P \) and a candidate lemma \( M \doteq N \) that is justified by \( \rlnm{Reduce} \). As the goal is assumed to be in normal form, we know that \( M \) is also in normal form. Thus there is a premise \( M \doteq N' \) where \( N \rightarrow_R^* N' \). We can apply this lemma directly, to leave the continuation \( C[N'\theta] \doteq P \). Of course, this is distinct from the continuation that we would arrive at if we used the unreduced lemma, i.e. \( C[N\theta] \doteq P \). However, if we normalise this original continuation to \( Q \doteq P' \), then, by confluence, the new continuation must also normalise to \( Q \doteq P' \), and we can proceed as normal. The comparison between these proofs can be seen in \cref{fig:Rewriting substitution}.
  \item \( \rlnm{Subst} \) If a lemma is itself justified by \( \rlnm{Subst} \), we can use the secondary lemma directly as contexts and substitutions are composable. Here we are observing that the order in which lemmas are applied is associative. However, choosing one of these as the canonical form, i.e. associating nested instances into the continuation, increases performance because the roles of the lemma and continuation are not symmetric --- we wish to reduce the number of choices for the former. This argument can also be seen in \cref{fig:Rewriting substitution}.
  \item Therefore, only those lemmas justified by \( \rlnm{Case} \) are considered for substitution.
\end{itemize}

In the proof that addition is commutative, for example, there are 16 vertices but only 3 instances of the \( \rlnm{Case} \), a significant reduction that mitigates the cost of verifying cycles.

\subsection{Verifying Cycles}\label{sec:Verifying cycles}

Our global condition on paths is undecidable in general.
If we restrict our attention to traces comprising variables and the substructural order, it becomes decidable.
Informally, this captures the space of typical proofs where induction concerns an explicit variable.

A comparable result was first shown by reduction to B{\"u}chi automata in the original work on cyclic proofs for first-order logic with inductive definitions~\cite{brotherston2005cyclic}.
Two \(\omega \)-regular languages are extracted from a preproof: the path language and the trace language.
It can then be checked whether the path language is included in the trace language, i.e.\ every path has an infinitely progressing trace.

Unfortunately, checking the inclusion of B{\"u}chi automata is doubly exponential in the number of vertices, as it involves complementing the automata~\cite{michel1988complementation}.
This procedure becomes onerous if several candidate proofs, the majority of which may be unsound, need to be checked throughout the proof search.
In the \textsc{Cyclist} theorem prover, soundness checking could take a significant proportion of the proof time~\cite{stratulat2019efficient}.

This approach to verifying cyclic proofs fails to take advantage of the incremental nature of the goal-orientated proof search, where proofs share a common prefix.
Furthermore, as soon as a cycle that does not satisfy the global condition is detected, there is no advantage to completing the proof.
Instead, we annotate the proof graph with an abstract domain representing the \(\omega \)-regular language of paths --- size-change graphs, originally developed for termination analysis~\cite{lee2001size}.
The workload is performed as each node is uncovered so that the soundness condition is represented explicitly.


\begin{definition}
  Let \( e(v) = \Gamma \vdash \phi \) and \( e(v') = \Gamma' \vdash \phi'  \) be two vertices in a preproof \( (V,\, e,\, r,\, p) \).
  A \defn{size-change graph} between \( v \) and \( v' \) is a labelled bipartite graph between \( \Gamma \) and \( \Gamma' \), i.e.\ a set of triples \( (x,\, y,\, l) \in \Gamma \times \Gamma' \times \{ \simeq, \lesssim \} \) where the labels mark equality or a decrease which are possible progress points.

  We write \( G : v \rightarrow v' \) for such a size-change graph, \( x \simeq y \in G \) if \( (x, \, y,\, l) \in G \) for any \( l \), and \( x \lesssim y \in G \) if, specifically, \( (x,\, y,\, \lesssim) \in G \).
  Labels from a simple lattice with \( {\lesssim} > {\simeq} \).
\end{definition}

\begin{definition}[Composition of size-change graphs]
  Given two size-change graphs \( G : v \rightarrow v' \) and \( G' : v' \rightarrow v'' \), then there is a size-change graph \( G' \circ G : v \rightarrow v'' \) defined as:
  \[
    G' \circ G \defeq \{ (x,\, z,\, l \sqcup l') \mid (x,\, y,\, l) \in G,\, (y,\, z,\, l') \in G' \}
  \]

  That is, there is an edge \( x \simeq z \) whenever there exists a variable \( y \) and edges \( x \simeq y \in G \) and \( y \simeq y \in G' \). It is decreasing if either edge is decreasing.
\end{definition}

The following definition associates a canonical size-change graph with each edge in a preproof.
Intuitively, an edge \( x \simeq y \in G_{(v,\, v')} \) indicates that \( x,\, y\) is a valid trace passing from \( v \) to \( v' \), and that it is a progress point if \( x \lesssim y \in G_{(v,\, v')} \).

\begin{definition}[The size-change graph of an edge]\label{def:Size-change edge}
  Let \( (V,\ e,\, r,\, p) \) be a preproof.
  For each edge \( (v,\, v') \in E \) of the underlying graph, \( G_{(v,\, v')} : v \rightarrow v' \) is defined as follows:

  \begin{itemize}
    \item If \( r(v) \) is an instance of \rlnm{Subst} with substitution \( \theta \) and \( v' \) is the lemma, then there is a non-decreasing edge \( x \simeq y \in G_{(v,\, v')} \) for all other variables such that \( x = \theta(y) \).
    \item If \( r(v) \) is an instance of \rlnm{Case} and the variable being analysed is \( x \), then there is a decreasing edge \(x \lesssim y \) for each fresh variable \( y \) introduced into \( v' \) and a non-decreasing edge \( z \simeq z \) for all variables.
    \item Otherwise, the size-change graph is simply the identity: \( z \simeq z \) for all variables in both environments.
  \end{itemize}
\end{definition}

The composition of these size-change graphs provides traces for general paths. And, by taking a generalisation of the transitive closure, we represent the space of possible infinite traces.

\begin{definition}
  The \defn{closure of a preproof} \( P = (V,\, e,\, r,\, p) \) is a set of size-change graphs, \( \mathrm{cl}(P) \), such that
  \begin{itemize}
    \item For each edge \( (v,\, v') \in E \), \( G_{(v,\, v')} \in \mathrm{cl}(P) \)
    \item If  \( G : v \rightarrow v' \in \mathrm{cl}(P) \) and \( G' : v' \rightarrow v'' \in \mathrm{cl}(P) \), then \( G' \circ G \in \mathrm{cl}(P) \)
  \end{itemize}
\end{definition}

\begin{lemma}\label{lem:Size-change trace}
  Suppose \( P = (V,\, e,\, r,\, p) \) is a preproof with a path \( v_0,\, \dots,\, v_n \) for some \( n > 0 \), then there is a size-change graph \( G : v_0 \rightarrow v_n \in \textrm{cl}(P) \) such that whenever \( x \simeq y \in G \) there is a trace \( x,\, \dots,\, y \) for this path, which has a progress point if \( x \lesssim y \in G \).
  \begin{appendixproof}
    Let us proceed by induction on the length of the path.
    \begin{itemize}
      \item If \( n = 0 \), then we are done.
      \item If \( n = 1 \), then \((v_0,\, v_1) \in E\). The size-change graph \( G_{(v_0,\, v_1)} \) from \cref{def:Size-change edge} is clearly sufficient.
      \item Suppose \( G : v_0 \rightarrow v_{n - 1} \) and \( (v_{n-1},\, v_n) \in E \). Then there is a size-change graph \( G_{(v_{n-1},\, v_n)} \). Consider the size-change graph \( G' = G_{(v_{n-1},\, v_n)} \circ G \). Suppose \( x \simeq y \in G' \), then \( x \simeq z \in G \) and \( z \simeq y \in G' \). By induction, there is a trace \( x,\, \dots,\, z \) for the path \( v_0,\, \dots,\, v_{n-1} \). It follows from \cref{def:Trace}, that \( x,\, \dots,\, z,\, y \) is a trace for the extended path to \( v_n \). Furthermore, we have a decreasing edge if either \( z \lesssim y \in G' \) or \( x \lesssim z \in G \). In either case, the trace clearly has a progress point.
    \end{itemize}
  \end{appendixproof}
\end{lemma}

As any infinite sequence of nodes is ultimately periodic, it is sufficient to look for decreasing edges in the size-change graphs representing cycles in the closure.

\begin{theorem}
  A cyclic preproof \( P = (V,\, e,\, r,\, p) \) is a proof if every \( G : v \rightarrow v \in \mathrm{cl}(P) \) such that \( G = G \circ G \) has a decreasing edge \( x \lesssim x \) for some variable \( x \).
  \begin{appendixproof}
    Let \( (v_i) \) be a path through \( P \).
    As \( V \) is finite, this path is ultimately periodic.
    That is, there exists some \( k \) and \( j \) such that \( v_{i + k + 1} = v_i \) for all \( i \geq j \).
    In which case, consider the corresponding size-change graph \( G : v_j \rightarrow v_j \in \mathrm{cl}(P) \) such that \( G = G \circ G \).
    By assumption, it has a decreasing edge \( x \lesssim x \) for some variable \( x \).
    It follows from \cref{lem:Size-change trace} that there is a trace \( x,\, \dots,\, x \) for the path \( v_j,\, \dots,\, v_{j + k} \), which has a progress point.
    Thus the infinite trace that cycles through this fintie trace satisfies the global condition for the path \( (v_i) \) as required.
  \end{appendixproof}
\end{theorem}

\section{Implementation and Empirical Evaluation}\label{sec:Implementation}

We implemented a prototype cyclic equational reasoning tool as a plugin for GHC 9.0.2 --- CycleQ.
It currently supports a small subset of Haskell, including top-level recursive functions, algebraic datatypes, and polymorphism.
The user adds equations to their program using the following syntax, and the plugin will attempt to prove them at compile-time, optionally outputting a cyclic proof graph if successful.

\begin{lstlisting}
  mapId :: List a -> Equation
  mapId xs = map id xs $\equiv$ xs
\end{lstlisting}

The tool performs a bounded depth-first search using the inference rules from \cref{fig:Inference rules}, in addition to a rule for function extensionality and decomposition of datatype constructors:
\[
  \prftree
  {\forall i \leq n}
  {M_i \doteq N_i}
  {k\ M_1\ \cdots\ M_n \doteq k\ N_1\ \cdots\ N_n}
\]
Although this rule is derivable from \( \rlnm{Subst} \), we distinguish it because it is not intended as a mechanism for creating cycles and can be applied eagerly, in a goal-directed manner, without incurring the cost associated with lemma generation.
Where more than one rule is applicable to a goal, the rules are prioritised as follows: reduction, reflexivity, congruence, function extensionality, subst, case analysis.
Once applied, the tool never backtracks past the first three as they always simplify the goal without loss of generality.
Furthermore, case analysis always selects a variable preventing further (non-strict) reduction, much like needed narrowing~\cite{antoy2000needed}.

\subsection{Evaluation}

There are very few implementations of cyclic proof systems, and their performance with equational goals is not well understood.
The \textsc{Cyclist} system~\cite{brotherston2012generic}, which is certainly the most developed, is known to have difficulty with equational reasoning and has issues with the verification of cycles~\cite{stratulat2019efficient}.
The primary objective of this evaluation is to demonstrate that our system, although simple, is reasonably efficient, avoiding a bottleneck in cycle verification.

We tested the tool against a standard benchmark suite of 85 induction problems concerning natural numbers, lists, and trees, originally used to test the IsaPlanner tool~\cite{dixon2003isaplanner}.
Since none of these concern mutual induction explicitly, we also designed a small number of problems around the representation of annotated, mutually recursive syntax trees, as shown in the introduction.
The results were obtained as an average of 10 runs on a 2.20GHz Intel\textregistered{} Core\texttrademark{} i5--5200U with 4 cores and
\SI{7.5}{GB} RAM.

\begin{figure}
  \includegraphics[width=0.47\textwidth]{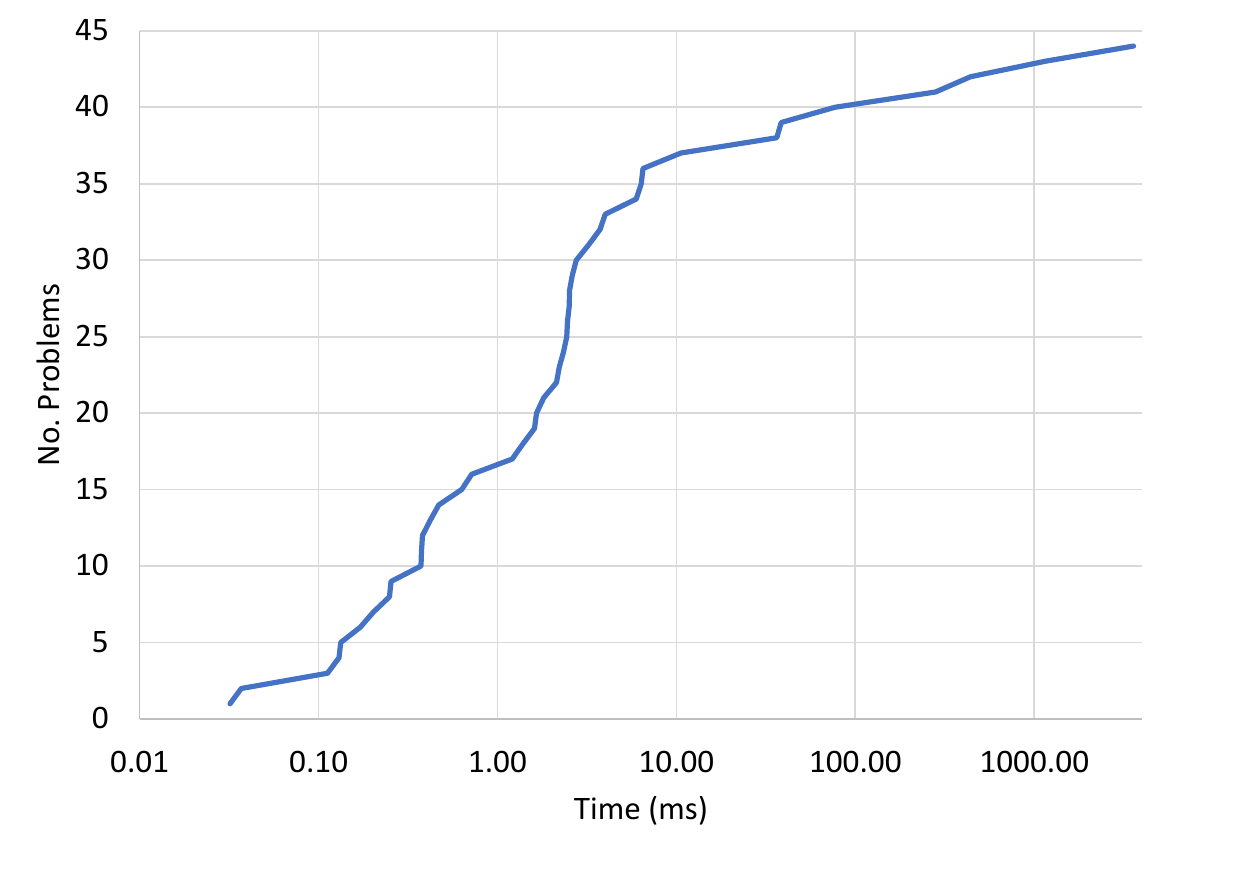}
  \caption{Summary of IsaPlanner benchmarks}\label{fig:Results}
\end{figure}

The number of IsaPlanner benchmark problems solved in a given time bound is plotted in \cref{fig:Results}.
The tool was able to solve 44 of the problems (13 were not in scope as they concerned conditional equations), with 40 of those solvable in under \SI{100}{ms}.  The average time for solvable IsaPlanner benchmarks was \SI{129}{ms}.
All the mutual induction problems were solved in \SI{5.3}{ms} on average.


\subsection{Limitations}

Although the tool performs efficiently on those 44 benchmark problems that it is able to solve, this number is relatively small.  By comparison: HipSpec proved 80, Zeno 82, CVC4 80, ACL2 74, Inductive Horn Clause Solving 68, IsaPlanner 47, and Dafny 45 (as reported by~\cite{claessen2013automating,unno2017automating}).

However, the following analysis shows that the problems CycleQ could not solve are attributable to two features that it lacks: \emph{conditional equations} and \emph{lemma discovery}, both of which are essentially orthogonal to cyclic reasoning.  Hence, we expect that our tool can incorporate these features in future work, which would allow for a more meaningful comparison.

First, many of the benchmark problems are themselves conditional equations.
Hence they simply fall outside of the scope of our system.

Some further 23 benchmarks, although not themselves conditional equations, require conditional equations internally in their proof.
Problem 4 is a typical example ---\\ \( \mathtt{S}\ (\mathtt{count}\ x\ xs) \doteq \mathtt{count}\ x\ (x : xs) \).
This can be solved by performing case analysis on the equality predicate, by assuming \( x == y \doteq \mathtt{False} \) or \( x == y \doteq \mathtt{True} \).
Our system does not currently have a mechanism for hypothetical reasoning of this form.
Since \( x \) may take infinitely many values in each case, and none of these enable a cycle, there is no other way to progress with the proof.

As far as we are aware, there is no reason why our system could not be extended in this direction (for example, by formulating the proof system using a judgement with an antecedent), but we felt it would overcomplicate the paper without adding any interesting new ideas.
These problems account for all those that Dafny solved that our tool did not.

The second reason some problems were unsolved is that they require lemmas, which was the case for the 4 remaining problems~\cite{rosenHipspecEval}.
Specifically, property 47 \emph{is} provable by our system when it is given the commutativity of \( \mathtt{max} \) and 54, 65, and 69 when given the commutativity of \( \mathtt{add} \).
Most comparable tools incorporate some form of lemma discovery, which is very powerful but orthogonal to this
work.
It is worth noting, however, that CycleQ solved a number of the benchmark problems designed to test strengthening and lemma discovery, despite not having a specialised tactic for either.
We look to incorporate a theory exploration system into our solver as future work, after which a direct comparison will be more insightful.

A couple of problems took significantly longer to solve.
And, unsurprisingly, these required the construction of larger proofs.
There are several factors to which this could be attributed: the branching factor of proof search, the increased number of lemmas, or the cost of verifying the global correctness condition.
In any case, we believe theory exploration could be used to mitigate this by allowing for smaller, more compositional proofs.

\section{Related Work and Conclusion}\label{sec:Related work}

As inductive definitions are ubiquitous in computer science, and functional programming in particular, a lot of work has been dedicated to developing tools that automate or aid equational reasoning over these structures.
However, as proof systems for inductive definitions don't admit cut elimination, most research is aimed at ``lemma'' discovery~\cite{bundy1999automation, johansson2019lemma}.

One technique for generating lemmas is to generalise the current goal by identifying common subterms, as implemented by ACL2~\cite{boyer2014computational}.
The heuristic was later refined by the Zeno tool that checks for counterexamples to prevent over generalisation, a common problem with the original method~\cite{sonnex2012zeno}.
It seems likely that the exploratory nature of cyclic proofs could be used to suggest generalisations from failed proofs without over-generalisation.

Proof planning was developed as a way to better control heuristics in automated reasoning tools~\cite{bundy1988use}.
It gave rise to Clam system and IsaPlanner~\cite{bundy1990s, dixon2003isaplanner, johansson2010dynamic}.
A lemma discovery strategy based on ``rippling'', a form of rewriting used in proof planning, was to construct a lemma from a failed proof~\cite{ireland1996productive}.
However, the required higher-order unification became a bottleneck to the technique's success~\cite{bundy1990s, dixon2003isaplanner, johansson2010dynamic}.

A radically different approach to lemma discovery is theory exploration~\cite{claessen2013automating}.
Instead of attempting to construct suitable lemmas analytically, theory exploration generates random lemmas and attempts to prove them in an incremental manner.
It is currently the state-of-the-art lemma discovery strategy, although it is hampered by scalability~\cite{johansson2019lemma}.

HipSpec is a tool that couples theory exploration with a traditional first-order theorem prover~\cite{claessen2010quickspec, rosen2012proving}.
As with the other approaches discussed so far, it ultimately relies on induction schema and thus cannot handle mutual induction.
We plan to integrate a theory exploration strategy into our tool, thus combining powerful lemma discovery with mutual induction.

The difficulties with induction has motivated a long line of work in inductionless induction~\cite{huet1982proofs,musser1980proving,kapur1987proof}.
While initially popular, as it can take advantage of general equational reasoning and rewriting techniques, the development of practical tools was limited~\cite{wirth2005history}.
However, the SPIKE theorem prover was based on this work and has since adopted a form of cyclic proof, but the relationship with our system is not completely clear~\cite{bouhoula1992spike, stratulat2020spike}.

Circular coinduction is a similar technique but for equations about coinductive structure~\cite{rocsu2009circular}.
Analogous to the ``expand'' operator in rewriting induction, the ``derivative'' of an equation is taken before it can be used as a hypothesis, where encodes the possible coinductive observations.

Although originally ill-suited to inductive theorem proving, many tools have successfully been built upon SMT solvers~\cite{suter2011satisfiability, leino2012automating, reynolds2015induction}.
More recently, induction has been incorporated into Horn clause solvers which, historically, struggle with domains such as non-linear arithmetic or some complex kinds of algebraic datatype~\cite{unno2017automating}.

Cyclic proofs have previously received attention for their application to program verification.
Specifically, a cyclic proof system for separation logic has been given that automatically verifies that a program terminates~\cite{brotherston2008cyclic, tellez2017automatically}.
Cyclic proof systems have recently been shown to subsume generic model-checking algorithms such as: lazy-abstraction with interpolants, property-directed reachability, and maximal conservativity for infinite game solving~\cite{tsukada2021software}.
As with the generic cyclic theorem prover \textsc{Cyclist}, it is the choice of ``matching-function'' or ``cut'' that determines exactly how the verification algorithm operates outside of the usual reasoning on the abstract domain.
Cyclic proofs have also been applied to program synthesis for pointer manipulating programs~\cite{itzhaky2021cyclic}.

The cost of verifying cycles has long been identified as a bottleneck of any tool based on cyclic proofs.
In Brotherston's thesis, he proposed an alternative approach --- ``trace manifolds''~\cite{brotherston2006sequent}.
A trace manifold is a set of trace segments that can be stitched together to construct a trace for any given path.
The trace segments are uniquely assigned, simplifying the space of traces significantly but excluding some complex cycles that might require different traces for the same path segment.
An alternative approach based on normalising the forms of cycles has been proposed and shown to be significantly more efficient~\cite{stratulat2019efficient, stratulat2021cyclist}.
The algorithm is polynomial.
However, there is no characterisation of exactly what patterns of cycles it is able to verify.


\begin{acks}                            
  We gratefully acknowledge the support of the \grantsponsor{EPSRC}{Engineering and Physical Sciences Research Council}{http://https://epsrc.ukri.org} (\grantnum{EPSRC}{EP/T006579/1}, \grantnum{EPSRC}{EP/T006595/1}) and the National Centre for Cyber Security via the UK Research Institute in Verified Trustworthy Software Systems.
\end{acks}

\bibliography{references,extrarefs}


\end{document}